\documentclass[11pt]{article}

\usepackage{geometry}
\usepackage[T1]{fontenc}
\usepackage[utf8]{inputenc}
\usepackage{amsmath, amssymb, amsthm}
\usepackage{authblk}

\theoremstyle{plain}
\newtheorem{theorem}{Theorem}
\newtheorem{lemma}[theorem]{Lemma}

\newtheorem{corollary}[theorem]{Corollary}
\theoremstyle{definition} %

\newtheorem{remark}[theorem]{Remark}
\newtheorem{definition}[theorem]{Definition}

\newenvironment{acknowledgements}{\section*{Acknowledgements}}{}

\usepackage{blkarray, bigstrut}
\usepackage{graphicx}
\usepackage{xspace}
\usepackage{colortbl}
\usepackage{multicol}
\usepackage{multirow}
\usepackage{enumerate}
\usepackage{tikz}
\usetikzlibrary{decorations.pathmorphing}
\usepackage{pgfplots}
\pgfplotsset{compat=1.16}
\usepackage{subcaption}
\usepackage{hyperref}

\makeatletter
\newif\if@restonecol
\makeatother

\usepackage[linesnumbered,ruled,vlined]{algorithm2e}
\let\oldnl\nl%
\newcommand{\nonl}{\renewcommand{\nl}{\let\nl\oldnl}}

  \DeclareMathOperator*{\join}{\Join}

\newcommand\topstrut[1][1.2ex]{\setlength\bigstrutjot{#1}{\bigstrut[t]}}
\newcommand\botstrut[1][0.9ex]{\setlength\bigstrutjot{#1}{\bigstrut[b]}}
\DeclareMathOperator{\sign}{sign}
\DeclareMathOperator{\Giv}{\mat{Giv}}
\newcommand{\R}{\mathbb{R}}
\newcommand{\tpl}[1]{\bar{#1}}
\newcommand{\mat}[1]{\mathbf{#1}}

\newcommand{\matspace}[2]{{\mathbb{R}}^{ #1 \times  #2}}
\newcommand {\calD}      {{\mathcal D}\xspace}
\newcommand {\calH}      {{\mathcal H}\xspace}
\newcommand {\calT}      {{\mathcal T}\xspace}
\newcommand{\norm}[1]{\ensuremath{||#1||_2}}
\newcommand{\pnorm}[2]{{\| {#1} \|}_{\mathrm{#2}}}

\newcommand{\bigO}{\mathcal{O}}

\newcommand{\frobnorm}[1]{{\| {#1} \| }_{\mathrm{F}}}
\newcommand{\matD}[1]{\mathbf{#1}}
\newcommand{\transpose}{\mathsf{T}}

\SetInd{0.1em}{0.5em}
\SetVlineSkip{0.12em}
\SetAlCapHSkip{0cm}
\SetAlgoSkip{}
\SetKwComment{Comment}{$\rhd$~}{}
\SetKwBlock{Begin}{}{}
\SetKw{Output}{return}
\SetKw{Choose}{choose}
\SetFuncSty{textsc}
\SetCommentSty{small}
\SetDataSty{textsc}
\SetArgSty{upshape}

\newcommand{\DS}[1]{\ensuremath{\mat{#1}}}%
\newcommand{\Data}{\DS{Data}\xspace}

\newcommand{\XScales}{\DS{scales}\xspace}

\newcommand{\Result}{\ensuremath{\mat{Out}}\xspace}

\newcommand{\Figaro}{\mbox{\FuncSty{FiGaRo}}\xspace}

\newcommand{\svdFigaro}{\mbox{\FuncSty{SVD-Fig}}\xspace}
\newcommand{\pcaFigaro}{\mbox{\FuncSty{PCA-Fig}}\xspace}
\newcommand{\pcaMKL}{\mbox{\FuncSty{PCA-MKL}}\xspace}

\newcommand{\svdMKL}{\mbox{\FuncSty{SVD-MKL}}\xspace}

\newcommand{\punto}{\hfill\ $\Box$}

\begin{document}

\newtheorem{insight}{Insight}

\title{Givens Rotations for QR Decomposition, SVD and PCA over Database Joins}

\author{Dan Olteanu}
\affil{University of Zurich, Switzerland. E-mail: olteanu@ifi.uzh.ch}
\author{Nils Vortmeier}
\affil{Ruhr University Bochum, Germany. E-mail: nils.vortmeier@rub.de}
\author{Đorđe Živanović}
\affil{University of Oxford, UK. E-mail: dorde.zivanovic@cs.ox.ac.uk}

\date{} 

\setcounter{Maxaffil}{2}
\renewcommand\Affilfont{\small}

\maketitle

\begin{abstract}
   This article introduces \Figaro, an algorithm for computing the upper-triangular matrix in the QR decomposition of the matrix defined by the natural join over relational data.
   \Figaro's main novelty is that it pushes the QR decomposition past the join. This leads to several desirable properties. For acyclic joins, it takes time linear in the database size and independent of the join size. Its execution is equivalent to the application of a sequence of Givens rotations proportional to the join size. Its number of rounding errors relative to the classical QR decomposition algorithms is on par with the database size relative to the join output size.

   The QR decomposition lies at the core of many linear algebra computations including the singular value decomposition (SVD) and the principal component ana\-lysis (PCA). We show how \Figaro can be used to compute the orthogonal matrix in the QR decomposition, the SVD and the PCA of the join output without the need to materialize the join output.

   A suite of experiments validate that \Figaro can outperform both in runtime performance and numerical accuracy the LAPACK library Intel MKL by a factor proportional to the gap between the sizes of the join output and input.

\end{abstract}

\section{Introduction}

This paper revisits the fundamental problem of computing the QR decomposition: Given a matrix $\mat A\in\mathbb{R}^{m\times n}$, its (thin) QR decomposition is the multiplication of the orthogonal matrix $\mat Q\in\mathbb{R}^{m\times n}$ and the upper-triangular matrix $\mat R\in\mathbb{R}^{n\times n}$~\cite{Higham}. This decomposition originated seven decades ago in works by Rutishauser~\cite{Rutishauser:QR:1958} and Francis~\cite{Francis:QR:1961}. There are three classical algorithms for QR decomposition: Gram-Schmidt and its modified version~\cite{Gram83,Schmidt07}, Householder~\cite{Householder58b}, and Givens rotations~\cite{Givens58}.

QR decomposition lies at the core of many linear algebra techniques and their machine learning applications~\cite{Stewart,matrix2016comp,SharmaPIM13} such as the matrix (pseudo) inverse and the least squares used in the closed-form solution for linear regression. In particular, the upper-triangular matrix $\mat R$ shares the same singular values with $\mat A$; its singular value decomposition can be used for the principal component analysis of $\mat A$; it constitutes a Cholesky decomposition of $\mat A^{\mat T}\mat A$, which is used for training (non)li\-near regression models using gradient descent~\cite{Schleich:F:2016}; the product of its diagonal entries equals (ignoring the sign) the determinant of $\mat A$, the product of the eigenvalues of $\mat A$, and the product of the singular values of $\mat A$.

In the classical linear algebra setting, the input matrix $\mat A$ is fully materialized and the process that constructs $\mat A$ is irrelevant. Our database setting is different: {\em $\mat A$ is not materialized and instead defined symbolically by the join of the relations in the input database. Our goal is to compute the QR decomposition of $\mat A$ without explicitly constructing $\mat A$.} This is desirable in case $\mat A$ is larger than the input database. By pushing the decomposition past the join down to the input relations, the runtime improvement is proportional to the size gap between the materialized join output and the input database.
This database setting has been used before for learning over relational data~\cite{Olteanu:Learning:2020}. Joins are used to construct  the training dataset from all available data sources and are unselective by design (the more labelled samples available for training, the better). This is not only the case for many-to-many joins; even key-fkey joins may lead to large (number of values in the) output, where data functionally determined by a key in a relation is duplicated in the join output as many times as this key appears in the join output. By avoiding the materialization of this large data matrix and by pushing the learning task past the joins, learning can be much faster than over the materialized matrix~\cite{Olteanu:Learning:2020}. Prior instances of this setting include pushing sum aggregates past joins~\cite{Larson,FAQ} and the computation of query probability in probabilistic databases~\cite{lazyeager-pdb}.

This article introduces \Figaro, an algorithm for computing the upper-triangular matrix $\mat R$ in the QR decomposition of the matrix $\mat A$ defined by the natural join of the relations in the input database.
The acronym \Figaro stands for {\bf Fa}ctorized {\bf Gi}vens {\bf Ro}tations with the letters {\bf a} and {\bf i} swapped. 

\Figaro enjoys several desirable properties.
\begin{enumerate}
  \item \Figaro's execution is equivalent to a sequence of Gi\-vens rotations over the join output. Yet instead of effecting the rotations individually as in the classical Givens QR decomposition, it performs fast block transformations whose effects are the same as for long sequences of rotations.

\item \Figaro takes time linear in the input data size and independently of the size of the (potentially much larger) join output for acyclic joins. It achieves this by pushing the computation past the joins. 

\item Its transformations can be applied independently and in parallel to disjoint blocks of rows and also to different columns. This sets it apart from inherently sequential mainstream methods for QR decomposition of materialized matrices such as Gram-Schmidt.

\item \Figaro can outperform both in runtime performance and accuracy the LAPACK library Intel MKL by a factor proportional to the gap between the join output and input sizes, which is up to three orders of magnitude in our experiments (Sec.\@~\ref{sec:experiments}). We show this to be the case for both key - foreign key joins over two real-world databases and for many-to-many joins of one real-world  and one  synthetic database.  We considered matrices with 2M-125M rows (84M-150M rows in the join matrix) and both narrow (up to 50 data columns) and wide (thousands of data columns). The choice of the join tree can significantly influence the performance (up to 47x). \Figaro is more accurate than  MKL as it introduces far less rounding errors in case the join output is larger than the input database.
\end{enumerate}

In Sec.\@~\ref{sec:beyondR}, we further extended \Figaro to compute: the orthogonal matrix $\mat Q$ in the QR decomposition of $\mat A$, the singular value decomposition (SVD) of $\mat A$, and the principal component analysis (PCA) of $\mat A$. These computations rely essentially on the fast and accurate computation of the upper-triangular matrix $\mat R$, are done {\em without materializing} $\mat A$, can benefit from pushing past joins dot products that are computed once and reused several times, and are amenable to partial computation in case only some output vectors are needed, such as the top-$k$ principal components. We show experimentally that these optimizations yield runtime and accuracy advantages of \Figaro over Intel MKL.

For QR decomposition, we designed an accuracy experiment of independent interest (Appendix~\ref{sec:accuracy-design}). The accuracy of an algorithm for QR decomposition is commonly based on how close the computed matrix $\mat Q$ is to an orthogonal matrix. We introduce an alternative approach that allows for a fragment $\mat R_\text{fixed}$ of the upper-triangular matrix $\mat R$ to be chosen arbitrarily and to serve as ground truth. Two relations are defined based on $\mat R_\text{fixed}$. We then compute the QR decomposition of the Cartesian product of the two relations and check how close $\mat R_\text{fixed}$ is to the corresponding fragment from the computed upper-triangular matrix. 

Our work complements prior work on linear algebra computation powered by database engines~\cite{LuoGGPJJ20,YuanJZTBJ21,Sagadeeva021,ElgoharyBHRR19,Boehm:SystemML:2014} and on languages that unify linear algebra and relational algebra~\cite{Suciu:LARADB:2017,Geerts:MATLANG:2021,DolmatovaAB20}. No prior work considered the interaction of QR decomposition with database joins. This requires a redesign of the decomposition algorithm from first principles. \Figaro is the first approach to take advantage of the structure and sparsity of relational data to improve the performance and accuracy of QR decomposition. The sparsity typically accounts for blocks of zeroes in the data. In our work, sparsity is more general as it accounts for blocks of arbitrary values that are repeated many times in the join output. \Figaro avoids such repetitions: It computes once over a block of values and reuses the computed result instead of recomputing it at every repetition.

We introduce \Figaro in several steps. We first explain how it works on the materialized join  (Sec.\@~\ref{sec:naive}) and then on the unmaterialized join modelled on any of its join trees (Sec.\@~\ref{sec:figaro}). \Figaro requires the computation of a batch of group-by counts over the join.  This can be done in only two passes over the input data (Sec.\@~\ref{sec:counts}). To obtain the desired upper-triangular matrix, \Figaro's output is further post-processed (Sec.\@~\ref{sec:postprocessing}).

This article extends a conference paper~\cite{Figaro:SIGMOD:2022} with three new contributions. 
\Figaro was extended to also compute: the orthogonal matrix $\mat Q$ in the QR decomposition of $\mat A$, an SVD of $\mat A$, and the PCA of $\mat A$ (Sec.\@~\ref{sec:beyondR}). This extension was implemented and benchmarked against mainstream approaches based on Intel MKL for both runtime performance and accuracy (Sec.\@~\ref{sec:experiments}). The article also extends the introductory example (Sec.\@~\ref{sec:intro-example}) and the related work (Sec.\@~\ref{sec:related}).

\subsection{Givens Rotations on the Cartesian Product}
\label{sec:intro-example}

We next showcase the main ideas behind \Figaro and start with introducing a (special case of) Givens rotation.
A $2 \times 2$ Givens rotation matrix is a matrix $\mat G = \begin{bmatrix}
\cos \theta & -\sin \theta \\
\sin \theta & \cos \theta
\end{bmatrix}$ for some angle $\theta$ (see Def.\@ \ref{def:givensrotation} for the definition of the general $d \times d$ case).
If $\theta$ is selected appropriately, applying a Givens rotation introduces zeros in matrices.
Consider a matrix $\mat B = \begin{bmatrix}
a \\
b
\end{bmatrix}$, where $a, b$ are real numbers with $b \neq 0$. We can visualize $\mat B$ as a vector in the two-dimensional Cartesian coordinate system. The matrix multiplication $\mat G \mat B$ represents the application of the rotation $\mat G$ to $\mat B$: its effect is to rotate the vector $\mat B$ counter-clockwise through the angle $\theta$ about the origin.
We can choose $\theta$ such that the rotated vector lies on the x-axis, so, having $0$ as second component:  
if we choose $\theta$ such that $\sin \theta = - \frac{\sign(a)b}{\sqrt{a^2 + b^2}}$ and $\cos \theta = \frac{|a|}{\sqrt{a^2 + b^2}}$~\cite{BindelDKM02}, then
$\mat G \mat B = \begin{bmatrix}
r \\
0
\end{bmatrix}$, where $r = \sign(a)\sqrt{a^2 + b^2}$.

When applied to matrices that represent the output of joins, Givens rotations can compute the QR decomposition more efficiently than for arbitrary matrices. We sketch this next for the Cartesian product of two unary relations.
Consider the matrices $\mat S = \begin{bmatrix}
s_1 \\
\cdots \\
s_p
\end{bmatrix}$, $\mat T = \begin{bmatrix}
t_1 \\
\cdots \\
t_q
\end{bmatrix}$, representing unary relations $S(Y_1)$ and $T(Y_2)$.
The matrix $\mat A = \begin{bmatrix}
\mat A_1 \\
\cdots \\
\mat A_p
\end{bmatrix}$, where $\mat A_i$ is the $q\times 2$  block matrix 
$\begin{bmatrix}
s_i & t_1 \\
\multicolumn{2}{c}{\cdots} \\
s_i & t_q
\end{bmatrix}$ for $i\in[p]$, represents the Cartesian product of these two unary relations, so, their natural join. 
We would like to compute the upper-triangular matrix $\mat R$ in the QR decomposition of $\mat A$: $\mat A = \mat Q \mat R$.\footnote{The matrix $\mat R$ does not depend on the order of rows in $\mat A$ respectively $\mat S$ and $\mat T$: Consider any permutation $\mat P$ of the rows in $\mat A$. Then, $\mat P \mat A = (\mat P \mat Q) \mat R$ so the permutation only affects the orthogonal matrix $\mat Q$ and not the upper-triangular matrix $\mat R$. When we go from relations to matrices, we can therefore order rows arbitrarily.}

The classical Givens rotations algorithm constructs the upper-triangular matrix $\mat R$ from $\mat A$ by using Givens rotations to zero each cell below the diagonal in $\mat A$. 
A sequence of Givens rotations can be used to set all entries below the diagonal of any matrix $\mat A$ to $0$, thus obtaining an upper triangular matrix. The left multiplication of these rotation matrices yields the orthogonal matrix $\mat Q$ in the QR decomposition of $\mat A$. This approach needs time quadratic in the input $\mat S$ and $\mat T$: it involves applying $2pq-3$ rotations, one rotation for zeroing each cell below the diagonal in $\mat A$.
This takes $13(2pq-3)$ multiplication, division, square, and square root operations. In contrast, \Figaro constructs $\mat R$ from $\mat S$ and $\mat T$ in linear time using $30(p+q)$ such operations.

Suppose we use Givens rotations to introduce zeros in the first column of the block  $\mat A_1 = \begin{bmatrix}
s_1 & t_1 \\
\multicolumn{2}{c}{\cdots} \\
s_1 & t_q
\end{bmatrix}$. So, we want to apply Givens rotations to a matrix column that consists of a single value and our goal is to set every occurrence of this value to $0$, apart of the first one.

\begin{figure}[t] \centering
\subfloat[Setting the entry of the second row to $0$.]{
\begin{tikzpicture}[yscale=-1]     
\node at (0,.8) {$\begin{bmatrix}
s \\ s \\ s
\end{bmatrix}$};
\draw[densely dashed] (0.4, 0.85) edge[bend left = 60, -stealth] (0.4, 0.45);

\begin{scope}[xshift=2.5cm, yshift=1.7cm, yscale=-1]
\begin{axis}[scale=0.5,
x=1cm, y=1cm,
  anchor=origin,
  axis lines=middle,
  every axis/.append style={font=\scriptsize},
  axis line style={-stealth,thick},
  xmin=-0.5,xmax=6,ymin=-0.5,ymax=4,
 ytick = \empty,
 xtick = \empty,
  title={},
      extra y ticks={3},
      extra y tick labels = {$s$},
      extra x ticks={3, 4.24},
      extra x tick labels = {$s$,  {\tiny $\sqrt{2}$}$s$},
      every x tick/.style={thick},
      every y tick/.style={thick}
        ]
\end{axis}
\draw[-stealth, thick] (0,0) -- (1.5,1.5);
\draw[-stealth, dashed] (44:2.125) arc (43.5:0:2.125);

\node at (3.5,0) {};
\end{scope}
\end{tikzpicture}
}
\quad
\subfloat[Setting the entry of the third row to $0$.]{
\begin{tikzpicture}[yscale=-1]   
\node at (0,.8) {$\begin{bmatrix}
\sqrt{2} s \\ 0 \\ s
\end{bmatrix}$};

\draw[densely dashed] (0.6, 1.2) edge[bend left = 60, -stealth] (0.6, 0.45);

\begin{scope}[xshift=2.5cm, yshift=1.7cm, yscale=-1]
\begin{axis}[scale=0.5,
x=1cm, y=1cm,
  anchor=origin,
  axis lines=middle,
  every axis/.append style={font=\scriptsize},
  axis line style={-stealth,thick},
  xmin=-0.5,xmax=6,ymin=-0.5,ymax=4,
 ytick = \empty,
 xtick = \empty,
  title={},
      extra y ticks={3},
      extra y tick labels = {$s$},
      extra x ticks={3, 4.24, 5.196},
      extra x tick labels = {$s$, {\tiny $\sqrt{2}$}$s$, {\tiny $\sqrt{3}$}$s$},
      every x tick/.style={thick},
      every y tick/.style={thick}
        ]
\end{axis}
\draw[-stealth, thick] (0,0) -- (2.11,1.5);
\draw[-stealth, dashed] (35:2.6) arc (34.5:0:2.6);
\node at (3,0) {};
\end{scope}
\end{tikzpicture}
}
\caption{Applying Givens rotations to a $1 \times 3$ matrix with all entries having the value $s$.}\label{figure:introexample}
\end{figure}
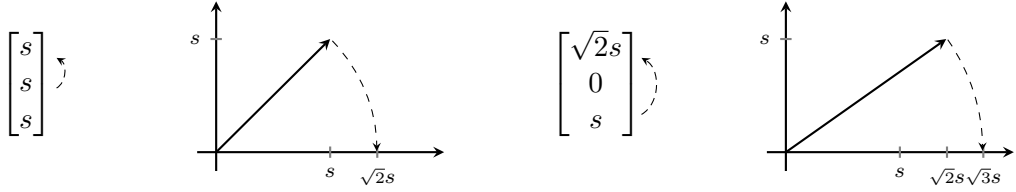

This setting is sketched in Figure~\ref{figure:introexample}: we have a matrix that consists of a single column and all entries are equal to the same value $s$. Using Givens rotations, all but the first entry need to be set to $0$.
The first rotation is applied to the first and the second occurrence of $s$, so to a vector that has the value $s$ in both components. This vector therefore has an angle of $45$ degrees to the $x$-axis, independent of the value of $s$. So, the rotation angle $\theta$ is independent of $s$. We see this also when we calculate $\sin \theta = - \frac{\sign(s)s}{\sqrt{s^2 + s^2}} = - \sqrt{\frac{1}{2}}$ and $\cos \theta = \frac{|s|}{\sqrt{s^2 + s^2}} = \sqrt{\frac{1}{2}}$ as explained above. The result of the rotation is that the second occurrence of $s$ is set to $0$, the first occurrence becomes $\sign(s) \sqrt{s^2 + s^2} = \sqrt{2}s$.

For the second rotation, we choose $\theta$ based on the value $\sqrt{2} s$ of the first entry and the value $s$ of the third entry of the first column, with the aim of setting the latter to $0$. The corresponding rotation with $\sin \theta =  - \sqrt{\frac{1}{3}}$ and $\cos \theta = \sqrt{\frac{2}{3}}$ sets the first entry to $\sqrt{3} s$. This argument can be continued for larger columns: the rotation, which is used to set the $k$-th occurrence of $s$ to $0$, has $\sin \theta =  - \sqrt{\frac{1}{k}}$ and $\cos \theta = \sqrt{\frac{k-1}{k}}$. 
We emphasize again that these rotations do not depend on the value $s$ but only on the number of such values.

So, to set for the block $\mat A_1 = \begin{bmatrix}
s_1 & t_1 \\
\multicolumn{2}{c}{\cdots} \\
s_1 & t_q
\end{bmatrix}$ all occurrences of the value $s_1$, apart the first, to $0$, we can use a series of Givens rotations with $\sin \theta = - \sqrt{\frac{1}{2}}, \ldots,  - \sqrt{\frac{1}{q}}$ and  $\cos \theta = \sqrt{\frac{1}{2}}, \ldots,  \sqrt{\frac{q-1}{q}}$, respectively.
The same applies to all other blocks $\mat A_i$ as well, so the same sequence of rotation angles can also be used to introduce zeros in the first column of every matrix $\mat A_i$.

\begin{insight}
For the matrix $\mat A_i$, there is a sequence $\mat G_s$ of Givens rotations that sets the first occurrence of $s_i$ to $s_i\sqrt{q}$ and all other occurrences to $0$. $\mat G_s$ is independent of the values in~$\mat A_i$.
\end{insight}

The blocks $\mat A_i$ have a second column, each consisting of the same values $t_1, \ldots, t_q$. As the same rotations are applied to each $\mat A_i$, the corresponding results $\mat A'_i = \begin{bmatrix}
s_i\sqrt{q} & t'_1 \\
0 & t'_2 \\
\multicolumn{2}{c}{\cdots} \\
0 & t'_q
\end{bmatrix}$
have the same values $t'_j$ throughout all blocks. They do not depend on the values $s_i$, so they can be computed from $\mat T$ alone. In Sec.\@~\ref{sec:rotations-products}, we show that this is possible in time $\bigO(q)$.

\begin{insight}
$\mat G_s$ yields the same values $t'_1,\ldots,t'_q$ in each matrix $\mat A'_i$. These values can be computed once.
\end{insight}

Among the blocks $\mat A'_i$, there are $p$ occurrences of the row  $\begin{bmatrix}
0 & t'_j
\end{bmatrix}$, for each $2 \leq j \leq q$, and $p$ rows of the form $\begin{bmatrix}
s_i\sqrt{q}\hspace*{.4em} & t'_1
\end{bmatrix}$. These rows can be grouped to have the same structure as the blocks $\mat A_i$ above: the second column consists of the same value $t'_j$. So, rotations analogous to $\mat G_s$ can be applied to these blocks to yield one row $\begin{bmatrix}
0 & t'_j \sqrt{p}
\end{bmatrix}$ and $p-1$ rows $\begin{bmatrix}
0 & 0
\end{bmatrix}$ for each $j\geq 2$, as well as one row $\begin{bmatrix}
s'_1\sqrt{q}\hspace*{.4em} & t'_1 \sqrt{p}
\end{bmatrix}$ and $p-1$ rows $\begin{bmatrix}
s'_i\sqrt{q}\hspace*{.4em} & 0
\end{bmatrix}$. The values $s'_i$ can be computed from $\mat S$ in time $\bigO(p)$ the same way as mentioned above and detailed in Sec.\@~\ref{sec:rotations-products}.

When all mentioned rotations are applied, we obtain the matrix
$\mat A''$ that consists of the blocks $\mat A''_1 = \begin{bmatrix}
s'_1\sqrt{q} & t'_1\sqrt{p} \\
0 & t'_2\sqrt{p} \\
\multicolumn{2}{c}{\cdots} \\
0 & t'_q\sqrt{p}
\end{bmatrix}$ and $\mat A''_i = \begin{bmatrix}
s'_i\sqrt{q} & 0 \\
0 & 0 \\
\multicolumn{2}{c}{\cdots} \\
0 & 0
\end{bmatrix}$ for $2\leq i \leq p$.

We observe the following four key points:
\begin{enumerate}[(1)]
\item The matrix $\mat A''$ has only $p+q$ non-zero values. We do not need to represent all zero rows as they are not part of the result $\mat R$.
\item The values in $\mat A''$ can be computed in one pass over $\mat S$ and $\mat T$. We need to compute: $\sqrt{p}$, $\sqrt{q}$, and the values $s'_1, \ldots, s'_p,t'_1,\ldots,t'_q$.
\item Our linear-time computation of the non-zero entries in $\mat A''$ yields the same result as quadratically many Givens rotations.
\item We see in Sec.\@~\ref{sec:rotations-products} that the computation of the values in $\mat A''$ does not require taking the squares of the input values, as done by the Givens QR decomposition. This means less rounding errors that would otherwise appear when squaring very large or very small input values.
\end{enumerate}

So far, $\mat A''$ is not upper-triangular. We obtain $\mat R$ from $\mat A''$ using a post-processing step that applies a sequence of $p+q-3$ rotations to zero all but the top three remaining non-zero values.

In summary, \Figaro needs $\bigO(p+q)$ time to compute $\mat R$ in our example.
Including post-processing, the number of needed squares, square roots, multiplications and divisions is at most $4(p+q)$, $3(p+q)$, $17(p+q)$ and $6(p+q)$, respectively.
In contrast, the Givens QR decomposition works on the materialized Cartesian product and needs $\bigO(pq)$ time and computes $4pq$ squares, $2pq$ square roots, $16pq$ multiplications and $4pq$ divisions.

The linear-time behaviour of \Figaro holds not only for the example of a Cartesian product, but for matrices defined by any acyclic join over arbitrary relations. In contrast, the runtime of any algorithm, which works on the materialized join output, as well as its square, square root, multiplication, and division computations are proportional to the size of the join output.

\section{\texorpdfstring{The \Figaro algorithm: Setup}{The Figaro algorithm: Setup}}
\label{sec:figaro-overview}

\textbf{Relations and Joins.}
A database $\calD$ consists of a set $S_1, \ldots, S_r$ of relations. Each relation has a schema $(Z_{1}, \ldots, Z_{k})$, which is a tuple of attribute names, and contains a set of tuples over this schema. We denote tuples $(Z_1, \ldots, Z_\ell)$ of attribute names by $\tpl Z$ and tuples $(z_1, \ldots, z_\ell)$ of values by $\tpl z$.
For every relation $S_i$, we denote by $\tpl X_i$ the tuple of its \emph{key} (join) attributes and by $\tpl Y_i$ the tuple of the \emph{data} (non-join) attributes. We allow key values of any type, e.g., categorical, while the data values are real numbers. We denote by $\tpl X_{ij}$ the join attributes common to both relations $S_i$ and $S_j$.
We consider the natural join of all relations in a database and write $\tpl X$ and $\tpl Y$ for the tuple of all key and data attributes in the join output. 
A database is \emph{fully reduced} if there is no dangling input tuple that does not contribute to the join output.
A join is ($\alpha$-)acyclic if and only if it admits a {\em join tree}~\cite{AbiteboulHV95}: In a join tree, each relation is a node and there is a path between the nodes for two relations $S_i$ and $S_j$ if all nodes along that path correspond to relations whose keys include $\tpl X_{ij}$.

\textbf{From Relations to Matrices.}
For a natural number $n$, we write $[n]$ for the set $\{1, \ldots, n\}$. Matrices are denoted by bold upper-case letters, (column) vectors by bold lower-case letters. Let a matrix $\mat A$ with $m$ rows and $n$ columns, a row index $i \in [m]$ and a column index $j \in [n]$. Then $\mat A[i : j]$ is the entry in row $i$ and column $j$, $\mat A[i :]$ is the $i$-th row, $\mat A[: j]$ is the $j$-th column of $\mat A$, and $|\mat A|$ is the number of rows in $\mat A$.
For sets $I \subseteq [m]$ and $J \subseteq [n]$, $\mat A[I : J]$ is the matrix consisting of the rows and columns of $\mat A$ indexed by $I$ and $J$, respectively.

Each relation $S_i$ is encoded by a matrix $\mat S_i$ that consists of all rows $(\tpl x_i, \tpl y_i)$ $\in S_i$. The relation representing the output of the natural join of the input relations is encoded by the matrix $\mat A$ whose rows $(\tpl x, \tpl y)$ are over the key attributes $\tpl X$ and data attributes $\tpl Y$.
We keep the keys and the column names as {\em contextual information}~\cite{DolmatovaAB20} to facilitate convenient reference to the data in the matrix.
{\em For ease of presentation, in this paper we use relations and matrices interchangeably following our above mapping. We use relational algebra over matrices to mean over the relations they encode.}

We use an indexing scheme for matrices that exploits their relational nature. We refer to the matrix columns by their corresponding attribute names. We refer to the matrix rows and blocks by tuples of key and data values. Consider for example a ternary relation $S$ with attribute names $X, Y_1, Y_2$. We represent this relation as a matrix $\mat S$ such that for every tuple $(x,y_1,y_2) \in S$ there is a matrix row $\mat S[x,y_1,y_2 :] = \begin{bmatrix}
x & y_1 & y_2
\end{bmatrix}$ with entries $\mat S[x,y_1,y_2 : X] = x$, $\mat S[x,y_1,y_2 : Y_1] = y_1$, and $\mat S[x,y_1,y_2 : Y_2] = y_2$.
We also use row indices that are different from the content of the row. 
We use $\ast$ to denote sets of indices. For example, $\mat S[x, \ast : X, Y_1]$ denotes the block that contains all rows of $\mat S$ whose identifier starts with $x$, restricted to the columns $X$ and $Y_1$.

\textbf{Input.} The input to \Figaro consists of (1) the set of matrices $\mat S_1,\ldots, \mat S_r$, one per relation in the input fully-reduced database $\calD$, and 
(2) a join tree $\tau$ of the acyclic natural join of these matrices (or equivalently of the underlying relations).\footnote{Arbitrary joins (including cyclic ones) can be supported by partial evaluation to acyclic joins: We construct a tree decomposition of the join and materialise its bags (sub-queries), cf.~prior works on learning over joins, e.g.,~\cite{Schleich:F:2016,Olteanu:FDB:2016,lmfao}. This partial evaluation incurs non-linear time. This is unavoidable under widely held conjectures.}

\textbf{Output.} 
\Figaro computes the upper-triangular matrix $\mat R$ in the QR decomposition of the matrix block $\mat A[: \tpl Y]$, which consists of the data columns $\tpl Y$ in the join of the database relations. It first computes an almost upper-triangular matrix $\mat R_0$ such that $\mat A[: \tpl Y] = \mat Q_0 \mat R_0$ for an orthogonal matrix $\mat Q_0$.  A post-processing step (Sec.\@~\ref{sec:postprocessing}) further decomposes $\mat R_0$: $\mat R_0 = \mat Q_1 \mat R$, where $\mat Q_1$ is orthogonal. Thus, $\mat A[: \tpl Y] = (\mat Q_0 \mat Q_1) \mat R$, where the final orthogonal matrix is $\mat Q = \mat Q_0 \mat Q_1$. \Figaro does not compute $\mat Q$ explicitly, it only computes $\mat R$.

The QR decomposition always exists; whenever $\mat A$ has full rank, and $\mat R$ has positive diagonal values, the decomposition is unique~\cite[Chapter~4, Theorem~1.1]{Stewart}.

\section{Heads and Tails}
\label{sec:rotations-products}

Given a matrix, the classical Givens QR decomposition algorithm repeatedly zeroes the values below its diagonal. Each zero is obtained by one Givens rotation. In case the matrix encodes a join output, a pattern emerges: The matrix consists of blocks representing the Cartesian products of arbitrary matrices and one-row matrices (see Fig.\@~\ref{fig:headtailsexample}(a)).
The effects of applying Givens rotations to such blocks are captured using so-called \emph{heads} and \emph{tails}. Before we introduce these notions, we recall the notion of Givens rotations.

\begin{definition}[Givens rotation]\label{def:givensrotation}
	A $d \times d$ \emph{Givens rotation matrix} is obtained from the $d$-dimensional identity matrix by changing four entries: $\mat G[i:i] = \mat G[j:j] = \cos \theta$, $\mat G[i:j] = - \sin \theta$ and $\mat G[j:i] = \sin \theta$, for some pair $i,j$ of indices and an angle $\theta$.
	\[\mat G =
 \begin{blockarray}{cccccccc}
                    &  &  & \scalebox{.75}{$i$} & & \scalebox{.75}{$j$} & &   \\[-1mm]
                  \begin{block}{c[ccccccc]}
                     & 1 & \cdots & 0 & \cdots & 0 & \cdots &  0 \topstrut\\
                     & \vdots & \ddots & \vdots &   &   \vdots &  & \vdots \\
            \scalebox{.75}{$i$\hspace{-1mm}} & 0 & \cdots & \cos \theta & \cdots  &   -\sin \theta & \cdots & 0 \\
            & \vdots &  & \vdots &  \ddots &   \vdots &  & \vdots \\
            \scalebox{.75}{$j$\hspace{-1mm}} & 0 & \cdots & \sin \theta & \cdots  &   \cos \theta & \cdots & 0 \\
            & \vdots &  & \vdots &   &   \vdots & \ddots & \vdots \\
            & 0 & \cdots & 0 & \cdots & 0 & \cdots &  1 \botstrut \\
                  \end{block}
                \end{blockarray}\]
 We denote such a $d$-dimensional rotation matrix for given $i$, $j$ and $\theta$ by $\Giv_d(i,j,\sin \theta, \cos \theta)$. \punto
\end{definition}

A rotation matrix is orthogonal.
The result $\mat B = \mat G \mat A$ of the product of a rotation $\mat G$ and a matrix $\mat A \in \R^{d \times d}$ is $\mat A$ except for the $i$-th and the $j$-th rows, which are subject to the counterclockwise rotation through the angle $\theta$ about the origin of the $2$-dimensional Cartesian coordinate system: for each column $c$, $\mbox{$\mat B[i:c]$} = \mbox{$\mat A[i:c]$} \cos \theta - \mbox{$\mat A[j:c]$} \sin \theta$ and $\mat B[j:c] = \mat A[i:c] \sin \theta + \mbox{$\mat A[j:c]$} \cos \theta$.

The following notions are used throughout the paper. Given matrices $\mat S \in \R^{m_1 \times n_1}$ and $\mat T \in \R^{m_2 \times n_2}$, their \emph{Cartesian product} $\mat S \times \mat T$ is the matrix $\begin{bmatrix}
    \mat A_1 \\ \vdots \\ \mat A_{m_1}
    \end{bmatrix}\in\R^{m_1m_2\times(n_1+n_2)}$ with $\mat A_i = \begin{bmatrix}
     \mat S[i:] & \mat T[1:] \\
     \vdots & \vdots \\
     \mat S[i:] & \mat T[m_2:] \\
    \end{bmatrix}$.
For $\mat S \in \R^{1 \times n}$ and $\mat v \in \R^{m}$, the \emph{Kronecker product} $\mat S \otimes \mat v$ is $\begin{bmatrix}
    v[1] \mat S \\
    \vdots \\
    v[m] \mat S
   \end{bmatrix}$.
We denote by $\norm{\mat v}$ the $\ell_2$ norm $\sqrt{\sum_{i=1}^m \mat v[i]^2}$ of a vector $\mat v \in \R^m$.

We now define the head and tail of a matrix, which we later use to express the Givens rotations over Cartesian products of matrices.

\begin{definition}[Head and Tail]
Let $\mat A$ be a matrix from $\R^{m \times n}$.

The \emph{head} $\calH(\mat A)\in \R^{1 \times n}$ of $\mat A$ is the one-row matrix \[\calH(\mat A) = \frac{1}{\sqrt{m}} \sum_{i = 1}^m \mat A[i:].\]

The \emph{tail} $\calT(\mat A)\in\R^{(m-1) \times n}$ of $\mat A$ is the matrix (for $j \in [m-1]$) \[\calT(\mat A)[j:] = \frac{1}{\sqrt{j+1}} \big(\sqrt{j} \mat A[j+1:] - \frac{\sum_{i=1}^j \mat A[i:]}{\sqrt{j}} \big) \hspace*{2em}\Box\]
\end{definition}

Let a matrix $\mat A$ that represents the Cartesian product of a one-row matrix $\mat S$ and an arbitrary matrix $\mat T$.
Then $\mat A$ is obtained by extending each row in $\mat T$ with the one-row $\mat S$, as in Fig.\@~\ref{fig:headtailsexample}(a). 
As exemplified in Sec.\@~\ref{sec:intro-example}, if all but the first occurrence of $\mat S$ in $\mat A$ is to be replaced by zeros using Givens rotations, the specific sequence of rotations only depends on the number of rows of $\mat T$ and not on the entries of $\mat S$. The result of these rotations can be described by the head and tail of $\mat T$.

\begin{lemma}\label{lem:cartesianproduct}
For matrices $\mat S \in \R^{1 \times n_1}$ and $\mat T \in \R^{m \times n_2}$, let $\mat A = \mat S \times \mat T \in \R^{m \times (n_1+n_2)}$ be their Cartesian product.
Let $\mat R_i = \Giv_m(1,i, -\frac{1}{\sqrt{i}}, \frac{\sqrt{i-1}}{\sqrt{i}})$, for all $i \in \{2, \ldots, m\}$, be a Givens rotation matrix and let $\mat G= \mat R_m \cdots \mat R_2$ be the orthogonal matrix that is the product of the rotations $\mat R_m$ to $\mat R_2$.

The matrix $\mat U = \mat G \mat A$ obtained by applying the rotations $\mat R_i$ to $\mat A$ is:
\[ \mat U =  \begin{bmatrix}
        \calH(\mat A) \\
        \calT(\mat A)
    \end{bmatrix} = \begin{bmatrix}
        \sqrt{m}\mat S & \calH(\mat T) \\
        \mat{0}^{{(m-1)} \times n_1} & \calT(\mat T)
    \end{bmatrix}.\]
\end{lemma}

In other words, the application of the $m-1$ rotations to $\mat A$ yields a matrix with four blocks: the head and tail of $\mat T$, $\mat S$ scaled by the square root of the number $m$ of rows in $\mat T$, and zeroes over $m-1$ rows and $n_1$ columns. So instead of applying the rotations, we may compute these blocks directly from $\mat S$ and $\mat T$.

\begin{figure}[t]
\centering
    \subfloat[Transformation of the Cartesian product of a one-row matrix $\mat S$ and an $m$-row matrix $\mat T$.]{
   $\mat A = \begin{bmatrix}
    \hspace*{.4em} \mat S & \multirow{4}{*}{\begin{tikzpicture} \fill[gray!20!white] (0,0) rectangle (.6,1.4); \node at (.3,.7){$\mat T$}; \end{tikzpicture} } \\
    \hspace*{.4em} \vdots \\
    \hspace*{.4em} \mat S 
    \end{bmatrix} 
    \quad \leadsto \quad 
    \begin{blockarray}{ccc} 
                     \begin{block}{c @{\hspace{0.2em}} [cc]}
    \calH(\mat A) = \hspace*{1em} & \sqrt{m} \mat S & \calH(\mat T) \topstrut \\
    \multirow{3}{*}{\begin{tikzpicture} \path[] (0,0) rectangle (.8,1); \node at (.4,.5){$\calT(\mat A) = \scalebox{0.9}{\Bigg\{}$}; \end{tikzpicture} } & 0 \vspace{-.7em} & \multirow{3}{*}{\begin{tikzpicture} \fill[gray!20!white] (0,0) rectangle (.8,1); \node at (.4,.5){$\calT(\mat T)$ }; \end{tikzpicture}}  \\
    & \vdots \vspace{-.2em} \\
    & 0 \\
   \end{block}\end{blockarray}$
   } \\ %
    \subfloat[Generalization of the transformation at (a), where $\mat S$ is scaled by a vector $\mat v = (v_1, \ldots, v_m)$ of positive real numbers.]{$\mat A = \begin{bmatrix}
    \hspace*{.4em} v_1 \mat S & \multirow{4}{*}{\begin{tikzpicture} \fill[gray!20!white] (0,0) rectangle (.6,1.4); \node at (.3,.7){$\mat T$}; \end{tikzpicture} }  \\
    \hspace*{.4em} \vdots  \\
    \hspace*{.4em} v_m \mat S 
    \end{bmatrix} \; \leadsto \hspace*{-.5em}
    \begin{blockarray}{ccc}
                     \begin{block}{c @{\hspace{0.2em}} [cc]}
    \calH(\mat A, \mat v) = \hspace*{1em} & \norm{\mat v} \mat S & \calH(\mat T, \mat v) \topstrut\\
    \multirow{3}{*}{\begin{tikzpicture} \path[] (0,0) rectangle (.8,1); \node at (.4,.5){$\calT(\mat A, \mat v) = \scalebox{0.9}{\Bigg\{}$}; \end{tikzpicture} } & 0 \vspace{-.7em} & \multirow{3}{*}{\begin{tikzpicture} \fill[gray!20!white] (0,0) rectangle (1.1,1); \node at (.55,.5){$\calT(\mat T, \mat v)$ }; \end{tikzpicture}}  \\
    & \vdots \vspace{-.2em} \\
    & 0 \\
   \end{block}
                   \end{blockarray}$}
   \caption{Visualization of Lemmas~\ref{lem:cartesianproduct} and \ref{lem:scaledcartesianproduct}: Matrices and results of applying the (generalized) head and tail. \label{fig:headtailsexample}}
   \end{figure}

We also encounter cases where blocks of matrices do not have the simple form depicted in Fig.\@~\ref{fig:headtailsexample}(a), but where the multiple copies of the one-row matrix $\mat S$ are scaled by different real numbers $v_1$ to $v_m$, as in  Fig.\@~\ref{fig:headtailsexample}(b). In these cases, we cannot compute heads and tails as in Lemma~\ref{lem:cartesianproduct} to capture the effect of a sequence of Givens rotations. 
However, a generalized version of heads and tails can describe these effects, as defined next.%

\begin{definition}[Generalized Head and Tail]\label{def:generalized-headtail}
For any vector $\mat{v} \in \R_{>0}^{m}$ and matrix $\mat A \in \R^{m \times n}$,
the \emph{generalized head} $\calH(\mat A, \mat{v})\in \R^{1 \times n}$ and  \emph{generalized tail} $\calT(\mat A, \mat{v})\in \R^{(m-1) \times n}$ of $\mat A$ weighted by $\mat v$ are:
\begin{align*}
    \calH(\mat A, \mat{v}) & = \frac{1}{\norm{\mat{v}}} \sum_{i = 1}^m \mat{v}[i] \mat A[i:]\\
    \calT(\mat A, \mat{v})[j:] & = \frac{1}{\norm{\mat{v}[1, \ldots, j+1]}} \\  \Big(\norm{\mat{v}[1, \ldots, j&]}  \mat A[j+1:] 
  - \frac{\mat{v}[j+1]\sum_{i=1}^j \mat{v}[i] \mat A[i:]}{\norm{\mat{v}[1, \ldots, j]}}  \Big) \hspace*{0.3em}\Box\
\end{align*}%
\end{definition}

If $\mat{v}$ is the vector of ones, then $\norm{\mat v} = \sqrt{m}$ and the generalized head and generalized tail become the head and tail, respectively.

We next generalize Lemma~\ref{lem:cartesianproduct}, where we consider each copy of $\mat S$ in $\mat A$ weighted by a positive scalar value from a weight vector $\mat v$: the $i$-th row of $\mat T$ is appended by $v_i \mat S$, for some positive real $v_i$, see Fig.\@~\ref{fig:headtailsexample}(b). This scaling is expressed using the Kronecker product~$\otimes$. Here again, to set all but the first (scaled) occurrences of $\mat S$ to zero using Givens rotations, we use that these rotations are independent of $\mat S$ and only depend on the scaling factors $\mat v$ and the number of rows in $\mat T$. We use $\calH(\mat A, \mat v)$ and $\calT(\mat A, \mat v)$ to construct the result of applying the rotations to $\mat A$.

\begin{lemma}\label{lem:scaledcartesianproduct}
Let $\mat{v} \in \R_{>0}^m, \mat S \in \R^{1 \times n_1}$, and $\mat T \in \R^{m \times n_2}$ be given and let $\mat A = \begin{bmatrix}
\mat S \otimes \mat{v} & \mat T
\end{bmatrix}\in \R^{m \times (n_1+n_2)}$.
Let $\mat R_i = \Giv_m(1,i, -\frac{\mat{v}[i]}{\norm{\mat{v}[1, \ldots, i]}}, \frac{\norm{\mat{v}[1, \ldots, i-1]}}{\norm{\mat{v}[1, \ldots, i]}})$, for all $i \in \{2, \ldots, m\}$, be a Givens rotation matrix and let $\mat G$ be the orthogonal matrix $\mat G = \mat R_m \cdots \mat R_2$.

The matrix $\mat U = \mat G \mat A$ obtained by applying the rotations $\mat R_i$ to $\mat A$ is:
\[ \mat U = \begin{bmatrix}
        \calH(\mat A, \mat v) \\
        \calT(\mat A, \mat v)
    \end{bmatrix} = \begin{bmatrix}
        \norm{\mat{v}}\mat S & \calH(\mat T,\mat{v}) \\
        \mat 0^{{(m-1)} \times n_1} & \calT(\mat T,\mat{v})
    \end{bmatrix}.\]
\end{lemma}

The generalized head and tail can be computed in linear time in the size of the input matrix $\mat A$. 

\begin{lemma}\label{lem:headtailcomplexity}
Given any matrix $\mat A \in \R^{m \times n}$ and vector $\mat v \in \R_{>0}^m$, the generalized head $\calH(\mat A, \mat v)$ and the generalized tail $\calT(\mat A, \mat v)$ can be computed in time $\bigO(mn)$.
\end{lemma}
The statement is obvious for generalized heads $\calH(\mat A, \mat v)$.
We give the explanation for the tail. 

Given the $\ell_2$-norm $\norm{\mat v[1, \ldots, j]}$ of the vector $\mat v$ restricted to the first $j$ entries, we can compute the same norm of $\mat v$, now restricted to the first $j+1$ entries, in constant time. Likewise, the sum $\sum_{i=1}^j \mat v[i] \mat A[i :]$ of the rows $\mat{v}[i] \mat A[i:]$, where $i$ goes up to $j$, can be reused to compute the sum of these rows up to $j+1$ in time proportional to the row length $n$. See Alg.\@~\ref{alg:tails} for details.
\begin{algorithm}[!tp]
  \DontPrintSemicolon%
  \FuncSty{GenTail}($\mat A$, $\mat v$)
\Begin{
    $\calT \gets \mat 0^{(m-1) \times n}$\\
    $\DataSty{sum\_A} \gets \mat 0^{1 \times n}$ \\
    $\DataSty{ssum\_v} \gets 0$ \\
     \For{$j = 1, \ldots, m-1$}{
      $\DataSty{ssum\_v} \gets \DataSty{ssum\_v} + \mat v[j]^2$ \\
      $\DataSty{sum\_A} \gets \DataSty{sum\_A} + \mat v[j] \mat A[j :]$ \\
     $\calT[j :] \gets \sqrt{\DataSty{ssum\_v}} \, \mat A[j+1 :] - \frac{\mat v[j+1] \DataSty{sum\_A}}{\sqrt{\DataSty{ssum\_v}}}$ \\
     $\calT[j :] \gets \frac{\calT[j :]}{\sqrt{\DataSty{ssum\_v} + \mat v[j+1]^2}}$
}
  \Output $\calT$
  }

  \caption{Computing generalized tails in linear time.}
  \label{alg:tails}
\end{algorithm}

The following lemma follows immediately from Def.\@~\ref{def:generalized-headtail}.
\begin{lemma}\label{lem:headtailproperties}
For any matrix $\mat A \in \R^{m \times n}$, any vector $\mat{v} \in R_{>0}^m$, and any numbers $k \in \R$ and $\ell \in \R_{>0}$, 
 the following holds:
\begin{align*}
 \calH(k \mat A, \ell\mat{v}) = &\  k\  \calH(\mat A, \mat{v}) \\
  \calT(k \mat A, \ell\mat{v}) = &\  k\  \calT(\mat A, \mat{v})
\end{align*}
\end{lemma}

\section{Example: Rotations on a join output}
\label{sec:naive}

Sec.\@~\ref{sec:intro-example} shows how to transform a Cartesian product of two matrices into an almost upper-triangular matrix by applying Givens rotations.  Sec.\@~\ref{sec:rotations-products} subsequently shows how the same result can be computed in linear time from the input matrices using the new operations head and tail, whose effects on matrices of a special form are summarized in Lemmas~\ref{lem:cartesianproduct} and~\ref{lem:scaledcartesianproduct}.

Towards an algorithm for the general case, this section gives an example of applying Givens rotations to a matrix that represents the natural join of four input relations. The insights gained from this example lead to the formulation of the \Figaro algorithm. 

Our approach is as follows. The result of joining two relations is a union of Cartesian products: for each join key $\tpl x$, all rows from the first relation that have $\tpl x$ as their value for the join attributes are paired with all rows from the second relation with $\tpl x$ as their join value. Sec.\@~\ref{sec:intro-example} and \ref{sec:rotations-products} explain how to process Cartesian products. Then we take the union of their results, so, accumulate the resulting rows in a single matrix.

\begin{figure}[t]
\centering
 \subfloat[Matrix $\mat S_1$]{\begin{tabular}{cc}
$X_{12}$  & $Y_1$ \\
\hline
$a_1$ &  $\mat S_1^{a_1}$ \\
$a_2$ &  $\mat S_1^{a_2}$ \\
\\[.4em]
\end{tabular}}
\qquad
 \subfloat[Matrix $\mat S_2$]{\begin{tabular}{cccc}
$X_{12}$ & $X_{23}$  & $X_{24}$  & $Y_2$ \\
\hline
$a_1$ & $b_1$ & $c_1$ & $\mat S_2^{a_1b_1c_1}$ \\
\multicolumn{4}{c}{$\cdots$} \\
$a_2$ & $b_2$ & $c_2$  & $\mat S_2^{a_2b_2c_2}$ 
\end{tabular}}
\qquad
 \subfloat[Matrix $\mat S_3$]{\begin{tabular}{cc}
$X_{23}$  & $Y_3$ \\
\hline
$b_1$ &  $\mat S_3^{b_1}$ \\
$b_2$ &  $\mat S_3^{b_2}$ \\
\\
\end{tabular}\:}

 \subfloat[Matrix $\mat S_4$]{\begin{tabular}{cc}
$X_{24}$  & $Y_4$ \\
\hline
$c_1$ &  $\mat S_4^{d_1}$ \\
$c_2$ &  $\mat S_4^{d_2}$
\end{tabular}\:}
\qquad
 \subfloat[Join tree $\tau$]{\hspace*{2em}\begin{tikzpicture}[inner sep=1pt]
\node (1) at (0,0) {$\mat S_1$};
\node (3) at (0,-0.5) {$\mat S_2$};
\node (4) at (-.3,-1) {$\mat S_3$};
\node (5) at (0.3,-1) {$\mat S_4$};
\draw (1) -- (3);
\draw (3) -- (4);
\draw (3) -- (5);
\end{tikzpicture}\hspace*{2em}}
\caption{Matrices and join tree used in Sec.\@~\ref{sec:naive}. \label{fig:examplenaive}}
\end{figure}
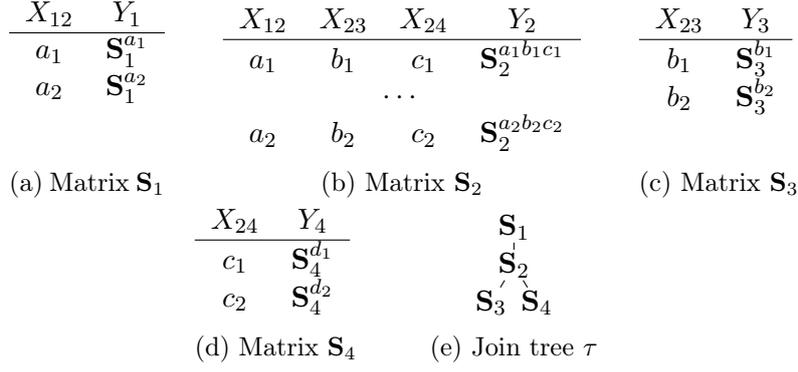

Figure~\ref{fig:examplenaive} sketches the input matrices $\mat S_1, \ldots, \mat S_4$ and the join tree $\tau$ used in this section. We explain $\mat S_1$, the other matrices are similar. 
The matrix $\mat S_1$ in Figure~\ref{fig:examplenaive}(a) has a column $X_{12}$ that represents the key attribute common with the matrix $\mat S_2$. 
It contains two distinct values $a_1$ and $a_2$ for its join attribute $X_{12}$. For each $a_j$, there is one vector $\mat S_1^{a_j}$ of values for the data column $Y_1$ in $\mat S_1$. The rows of $\mat S_1$ are the union of the Cartesian products $[a_j] \times \mat S_1^{a_j}$, for all $j \in \{1,2\}$.

The  output $\mat A$ of the natural join of $\mat S_1, \ldots, \mat S_4$ contains the Cartesian product
\[[a_j \;  b_k \;  c_\ell] \times \mat S_1^{a_j} \times \mat S_2^{a_ j b_k c_\ell} \times \mat S_3^{b_k} \times \mat S_4^{c_\ell}\]
for every tuple $(a_j, b_k, c_\ell)$ of join keys, with $j,k,l \in \{1,2\}$.
As mentioned in Sec.\@~\ref{sec:figaro-overview}, we want to transform the matrix $\mat A[\tpl Y]$, i.e., the projection of $\mat A$ to the data columns $\tpl Y$, into a matrix $\mat R_0$ that is almost upper-triangular. In particular, the number of rows with non-zero entries in $\mat R_0$ needs to be linear in the size of the input matrices. We do this by repeatedly identifying parts of the matrix $\mat A[\tpl Y]$ that have the form  depicted in Figure~\ref{fig:headtailsexample} and by applying Lemmas~\ref{lem:cartesianproduct} and~\ref{lem:scaledcartesianproduct} to these parts. Recall that each of these applications corresponds to the application of a sequence of Givens rotations.
 
The matrix $\mat A[: \tpl Y]$ contains the rows with the following Cartesian product associated with the join keys $(a_1, b_1, c_1)$:
\begin{center}
\begin{tabular}{ccccccc}
$Y_1$ & & $Y_2$ & & $Y_3$ & & $Y_4$   \\
\hline
$\mat S_1^{a_1}$ & $\times$  & $\mat S_2^{a_1b_1c_1}$ & $\times$ & $\mat S_3^{b_1}$& $\times$ & $\mat S_4^{c_1}$
\end{tabular}
\end{center}

In turn, this Cartesian product is the union of Cartesian products $[t_1 \; t_2 \; t_3] \times \mat S_4^{c_1}$, for each choice $t_1, t_2, t_3$ of rows of $\mat S_1^{a_1}$, $\mat S_2^{a_1b_1c_1}$ and respectively $\mat S_3^{b_1}$. The latter products have the form depicted in Figure~\ref{fig:headtailsexample}(a): the Cartesian product of a one-row matrix $[t_1 \; t_2 \; t_3]$ and an arbitrary matrix $\mat S_4^{c_1}$. By applying Lemma~\ref{lem:cartesianproduct} to each of them, we obtain the following rows, where we use  $\alpha = \sqrt{|\mat S_4^{c_1}|}$:

\begin{center}
\begin{tabular}{ccccccccc}
$Y_1$ & & $Y_2$ & & $Y_3$ & & $Y_4$  \\
\hline
$\alpha \mat S_1^{a_1}$ & $\times$ & $\alpha \mat S_2^{a_1b_1c_1}$ & $\times$ & $\alpha \mat S_3^{b_1}$ & $\times$  & $\calH(\mat S_4^{c_1})$ \\
$0$ & & $0$ &  & $0$ &  &$\calT(\mat S_4^{c_1})$ \\
$\vdots$ & & $\vdots$ &  & $\vdots$ &  & $\vdots$ \\
$0$ & & $0$ & & $0$ &  & $\calT(\mat S_4^{c_1})$
\end{tabular}
\end{center}

This matrix contains one copy of $\calT(\mat S_4^{c_1})$ for every choice of $t_1, t_2, t_3$. 
More copies of $\calT(\mat S_4^{c_1})$ arise when we analogously apply Lemma~\ref{lem:cartesianproduct} to the rows of $\mat A[: \tpl Y]$ associated with join keys $(a_j, b_k, c_1)$ that are different from $(a_1, b_1, c_1)$. 
In total, the number of copies of $\calT(\mat S_4^{c_1})$ we obtain is \[\sum_{j,k \in \{1,2\}} |\mat S_1^{a_j}|\cdot|\mat S_2^{a_jb_kc_1}|\cdot|\mat S_3^{b_k}| = \big| \sigma_{X_{24}=c_1} (\mat S_1 \join \mat S_2 \join \mat S_3) \big|,\]
so, the size of the join of all matrices except  $\mat S_4$, where the attribute $X_{24}$ is fixed to $c_1$. We denote this number by $\Phi_4^\circ(c_1)$. %

We apply more Givens rotations to the rows of the form $[0 \; 0 \; 0] \times \mat \calT(\mat S_4^{c_1})$. These are again Cartesian products, namely the Cartesian products of the zero matrix of dimension $\Phi_4^\circ(c_1) \times 3$ and each row $t$ of $\calT(S_4^{c_1})$. We can apply again Lemma~\ref{lem:cartesianproduct} to these Cartesian products and obtain the rows
\begin{center}
\begin{tabular}{ccccc}
$Y_1$ & $Y_2$ & $Y_3$ & $Y_4$  \\
\hline
$0$ & $0$ & $0$   & $\sqrt{\Phi^\circ_4(c_1)} \calT(\mat S_4^{c_1})$ \\
$0$ & $0$ & $0$   & $0$ \\
$\vdots$ & $\vdots$ & $\vdots$   & $\vdots$ \\
$0$ & $0$ & $0$ & $0$. 
\end{tabular}
\end{center}

The rows that only consist of zeros can be discarded, rows with non-zero entries become part of the (almost upper-triangular) result matrix $\mat R_0$.

We now turn back to the remaining rows, which are of the form

\begin{center}
\begin{tabular}{ccccccccccc}
$Y_1$ & & $Y_2$ & & $Y_3$ & & $Y_4$  \\
\hline
$\alpha \mat S_1^{a_1}$ & $\times$  & $\alpha \mat S_2^{a_1b_1c_1}$ & $\times$ & $\alpha \mat S_3^{b_1}$ & $\times$ & $\calH(\mat S_4^{c_1})$. 
\end{tabular}
\end{center}

As above, we view these rows as unions of (column-permutations of) Cartesian products $[\alpha t_1 \; \alpha t_2  \; \calH(\mat S_4^{c_1})] \times \alpha S_3^{b_1}$, for each choice $t_1, t_2$ of rows of $\mat S_1^{a_1}$ and $\mat S_2^{a_1b_1c_1}$, respectively.
The corresponding applications of Lemma~\ref{lem:cartesianproduct} yield the rows

\begin{center}
\setlength{\tabcolsep}{5.5pt}
\begin{tabular}{ccccccccccc}
$Y_1$ & & $Y_2$ & & $Y_3$ & & $Y_4$  \\
\hline
$\gamma$ $\mat S_1^{a_1}$ & $\times$  & $\gamma$ $\mat S_2^{a_1b_1c_1}$ & $\times$ & $\alpha \calH(\mat S_3^{b_1})$ & $\times$ & $\beta \calH(\mat S_4^{c_1})$ \\
$0$ & & $0$ &  & $\alpha \calT(\mat S_3^{b_1})$ &  &$0$ \\
$\vdots$ & & $\vdots$ &  & $\vdots$ &  & $\vdots$ \\
$0$ & & $0$ &  & $\alpha \calT(\mat S_3^{b_1})$ &  & $0$
\end{tabular}
\end{center}
where $\beta = \sqrt{|\mat S_3^{b_1}|}$ and $\gamma = \alpha \beta$.
Considering also the rows that we analogously obtain for join keys $(a_j, b_1, c_\ell)$ different from $(a_1, b_1, c_1)$, we get $\sum_{j\in \{1,2\}} |\mat S_1^{a_j}|\cdot|\mat S_2^{a_jb_1c_1}|$ copies of $\sqrt{|\mat S_4^{c_1}|}\calT(\mat S_3^{b_1})$ in total, and $\sum_{j \in \{1,2\}} |\mat S_1^{a_j}|\cdot|\mat S_2^{a_jb_1c_2}|$ copies of $\sqrt{|\mat S_4^{c_2}|}\calT(\mat S_3^{b_1})$.

We apply further Givens rotations to the rows that consist of zeros and the scaled copies of $\calT(\mat S_3^{b_1})$. For each row $t$ of $\calT(\mat S_3^{b_1})$, we apply this time Lemma~\ref{lem:scaledcartesianproduct} to the matrix of all scaled copies of $t$ and three columns of zeros. The result contains some all-zero rows and one row with entry $t$ scaled by the $\ell_2$ norm of the vector of the original scaling factors.
This factor is the square root of
\begin{align*}
 & \sum_{j \in \{1,2\}} |\mat S_1^{a_j}|\cdot|\mat S_2^{a_jb_1c_1}|\cdot \sqrt{|\mat S_4^{c_1}|}^2 %
  + \sum_{j \in \{1,2\}} |\mat S_1^{a_j}|\cdot|\mat S_2^{a_jb_1c_2}|\cdot \sqrt{|\mat S_4^{c_2}|}^2 \\
  = & \sum_{j,\ell \in \{1,2\}} |\mat S_1^{a_j}|\cdot|\mat S_2^{a_jb_1c_\ell}|\cdot|\mat S_4^{c_\ell}| \\
  = & \: \big| \sigma_{X_{23}=b_1} (\mat S_1 \join \mat S_2 \join \mat S_4) \big|,
\end{align*} 
which, similarly as above, is the size of the join of all matrices except  $\mat S_3$, where the attribute $X_{23}$ is fixed to $b_1$. We denote this number by $\Phi_3^\circ(b_1)$.
In summary, the result of applying Lemma~\ref{lem:scaledcartesianproduct} contains, besides some all-zero rows, the rows

\begin{center}
\begin{tabular}{ccccc}
$Y_1$ & $Y_2$ & $Y_3$ & $Y_4$  \\
\hline
$0$ & $0$ & $\sqrt{\Phi^\circ_3(b_1)}$ $\mat \calT(\mat S_3^{b_1})$   & $0$.
\end{tabular}
\end{center}

The remaining columns are processed analogously.
Up to now, the transformations result in all-zero rows, rows that contain the scaled tail of a part of an input matrix with other columns being $0$, and one row for every join key $(a_j, b_k, c_\ell)$, for $j,k,\ell \in \{1,2\}$. We give the resulting rows for $a_j = a_1$, where for $k,\ell \in \{1,2\}$ we use $\alpha_{k \ell} = \sqrt{|\mat S_1^{a_1}|\cdot|\mat S_2^{a_1 b_k c_\ell}|\cdot|\mat S_3^{b_k}|\cdot|\mat S_4^{c_\ell}|}$.

\begin{center}
\begin{tabular}{cccccc}
$Y_1$ & $Y_2$ & $Y_3$ & $Y_4$  \\
\hline
\small $\frac{\alpha_{11}}{\sqrt{|\mat S_1^{a_1}|}} \calH(\mat S_1^{a_1})$  &  \small $\frac{\alpha_{11}}{\sqrt{|\mat S_2^{a_1 b_1 c_1}|}} \calH(\mat S_2^{a_1 b_1 c_1}) $  &  \small $\frac{\alpha_{11}}{\sqrt{|\mat S_3^{b_1}|}} \calH(\mat S_3^{b_1})$  &  \small $\frac{\alpha_{11}}{\sqrt{|\mat S_4^{c_1}|}} \calH(\mat S_4^{c_1})$     \\
\multicolumn{4}{c}{$\vdots$} \\[0.2em]
\small $\frac{\alpha_{22}}{\sqrt{|\mat S_1^{a_1}|}} \calH(\mat S_1^{a_1})$  &  \small $\frac{\alpha_{22}}{\sqrt{|\mat S_2^{a_1 b_2 c_2}|}} \calH(\mat S_2^{a_1 b_2 c_2}) $  &  \small $\frac{\alpha_{22}}{\sqrt{|\mat S_3^{b_2}|}} \calH(\mat S_3^{b_2})$  &  \small $\frac{\alpha_{22}}{\sqrt{|\mat S_4^{c_2}|}} \calH(\mat S_4^{c_2})$
\end{tabular}
\end{center}

We observe that the entries in the column $Y_1$ only differ by the scaling factors $\alpha_{k \ell}$. 
It follows that we can apply Lemma~\ref{lem:scaledcartesianproduct}, with $\mat S = \calH(\mat S_1^{a_1})$, $\mat T$ consisting of the columns $Y_2, Y_3, Y_4$ above, and 
$\mat v =
\begin{bmatrix}
  \frac{\alpha_{11}}{\sqrt{|\mat S_1^{a_1}|}} \\
  \vdots \\
  \frac{\alpha_{22}}{\sqrt{|\mat S_1^{a_1}|}}
 \end{bmatrix}
=
\begin{bmatrix}
  \frac{\sqrt{|\mat S_1^{a_1}|\cdot|\mat S_2^{a_1 b_1 c_1}|\cdot|\mat S_3^{b_1}|\cdot|\mat S_4^{c_1}|}}{\sqrt{|\mat S_1^{a_1}|}} \\
  \vdots \\
  \frac{\sqrt{|\mat S_1^{a_1}|\cdot|\mat S_2^{a_1 b_2 c_2}|\cdot|\mat S_3^{b_2}|\cdot|\mat S_4^{c_2}|}}{\sqrt{|\mat S_1^{a_1}|}}
 \end{bmatrix}
=
\begin{bmatrix}
 \sqrt{|\mat S_2^{a_1b_1c_1}| \scalebox{1}{$\cdot$} |\mat S_3^{b_1}| \scalebox{1}{$\cdot$} |\mat S_4^{c_1}|} \\
 \vdots \\
 \sqrt{|\mat S_2^{a_1b_2c_2}| \scalebox{1}{$\cdot$} |\mat S_3^{b_2}| \scalebox{1}{$\cdot$} |\mat S_4^{c_2}|}
\end{bmatrix}$.

The first row of the result is the generalized head of the above rows. For column $Y_1$, this is $\norm{\mat{v}} \calH(\mat S_1^{a_1})$, where $\norm{\mat{v}} = \sqrt{\sum_{k, \ell \in \{1,2\}}|\mat S_2^{a_1b_kc_\ell}|\cdot|\mat S_3^{b_k}|\cdot|\mat S_4^{c_\ell}|} = \sqrt{|\sigma_{X_{12}=a_1} \mat S_2 \join \mat S_3 \join \mat S_4|}$. The term under the square root is the size of the join of all matrices in the subtree of $\mat S_3$ in $\tau$, where $X_{12}$ is fixed to $a_1$. We denote this number by $\Phi_3^\downarrow(a_1)$. We skip the discussion for the other columns.

The remaining rows are formed by the generalized tail. Column $Y_1$ has only zeros. For column $Y_4$, the result is $\sqrt{|\mat S_1^{a_1}|} \mat T_{4,a_1}$, where $\mat T_{4,a_1} = \calT(\begin{bmatrix}
\sqrt{|\mat S_2^{a_1 b_1 c_1}|\cdot|\mat S_3^{b_1}|} \calH(\mat S_4^{c_1})  \\
\vdots \\
\sqrt{|\mat S_2^{a_1 b_2 c_2}|\cdot|\mat S_3^{b_2}|} \calH(\mat S_4^{c_2})
\end{bmatrix},\mat v)$, and analogously for the columns $Y_2$ and $Y_3$.

\begin{remark}
Suppose $\mat S_1$ would have another child $\mat S_5$ in $\tau$, so, $\mat S_1$ would also have the join attributes $X_{15}$. Then we would get one scaled copy of $\mat T_{4,a_1}$ for every value of $X_{15}$ that occurs together with the value $a_1$ for $X_{12}$. We  would then apply Lemma~\ref{lem:scaledcartesianproduct} again. The scaling factor for $T_{4,a_1}$ we obtain in the general case is the square root of the size of the join of all relations that are \emph{not} in the subtree of $\mat S_3$ in $\tau$, where $X_{12}$ is fixed to $a_1$. 
In the simple case considered here, this value is $\sqrt{|\mat S_1^{a_1}|}$.
\end{remark}%

To sum up, the resulting non-zero rows are of the following form:
\begin{enumerate}
 \item $\sqrt{\Phi^\circ_i(\tpl x_i)} \calT(\mat S_i^{\tpl x_i})$, for all $1 \leq i \leq 4$ and values $\tpl x_i$ for the join attributes of $\mat S_i$,
 \item $\sqrt{|\mat S_1^{a_j}|} [\mat T_{2,a_j} \; \mat T_{3,a_j} \; \mat T_{4,a_j}]$, for $j \in \{1,2\}$, and
 \item one row of scaled generalized heads for $a_1$ and $a_2$. \quad
\end{enumerate}

The number of rows of type (1) is bounded by the overall number of rows in the input matrices minus the number of different join keys, as the tail of a matrix has one row less than the input matrix. The number of rows of type (2) is bounded by the number of different values for the join attributes $(X_{12},X_{23}, X_{24})$ minus the number of different values for $X_{12}$. The number of rows of type (3) is bounded by the number of different values for the join attribute $X_{12}$. 
Together, the number of non-zero rows is bounded by the overall number of rows in the input matrices, as desired. 

\section{Scaling Factors as Count Queries}
\label{sec:counts}

The example in Sec.\@~\ref{sec:intro-example} applies Givens rotations to a Cartesian product. The resulting matrix uses scaling factors that are square roots of the numbers of rows of the input matrices. These numbers can be trivially computed directly from these matrices. In case of a matrix defined by joins, as seen in the preceding Sec.\@~\ref{sec:naive}, these scaling factors are defined by group-by count queries over the joins. \Figaro needs three types of such count queries at each node in the join tree. They can be computed together in linear time in the size of the input matrices. This section defines these queries and explains how to compute them efficiently.

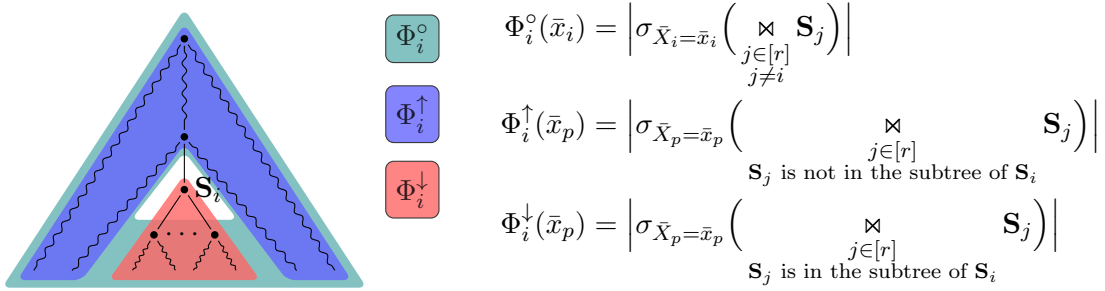
\begin{figure}[t]
\centering
\begin{subfigure}[b]{0.4\textwidth}
\begin{tikzpicture}
 \path[fill = teal!60!white, opacity = 0.8, rounded corners=3pt] (0,2.4) -- (-2.4,-1.3) -- (2.4,-1.3) -- cycle;
 \path[fill = white, opacity = 1, rounded corners=3pt] (0,.5) -- (-.7,-.4) -- (.7,-.4) -- cycle;
 \path[fill = red!60!white, opacity = 0.8, rounded corners=3pt] (0,0.2) -- (-1,-1.2) --  (1,-1.2) -- cycle;
 \path[fill = blue!60!white, opacity = .8, rounded corners=3pt] (0,2.2) -- (-2.2,-1.2) --  (-1.35,-1.2) -- (0,.6) -- (1.35,-1.2) -- (2.2,-1.2) -- cycle;
\node[inner sep=1pt, circle, fill=black] (r) at (0,2) {};

\node[inner sep=1pt, circle, fill=black, label={[label distance=0mm]0:}] (p) at (0,.7) {};

\node[inner sep=1pt, circle, fill=black, label={[label distance=-.6mm]0:$\mat S_i$}] (i) at (0,0) {};

\node[inner sep=1pt, circle, fill=black] (j1) at (-.4,-.6) {};
\node[inner sep=1pt, circle, fill=black] (j2) at (.4,-.6) {};
\node at (0,-.6) {$\cdots$};
\path[shorten >=1pt,shorten <=1pt]
 (r) edge[decorate,decoration={snake,amplitude=.3mm,segment length=2mm,pre length=3pt, post length=3pt}] (p)
 (r) edge[decorate,decoration={snake,amplitude=.3mm,segment length=2mm,pre length=3pt, post length=1pt}] (-2,-1.1)
 (r) edge[decorate,decoration={snake,amplitude=.3mm,segment length=2mm,pre length=3pt, post length=1pt}] (2,-1.1)
 (p) edge[decorate,decoration={snake,amplitude=.3mm,segment length=2mm,pre length=3pt, post length=1pt}] (-1.5,-1.1)
 (p) edge[decorate,decoration={snake,amplitude=.3mm,segment length=2mm,pre length=3pt, post length=1pt}] (1.5,-1.1)
 (p) edge (i)
 (i) edge (j1)
 (i) edge (j2)
 (j1) edge[decorate,decoration={snake,amplitude=.2mm,segment length=1mm,pre length=3pt, post length=1pt}] (-.7,-1.1)
 (j1) edge[decorate,decoration={snake,amplitude=.2mm,segment length=1mm,pre length=3pt, post length=1pt}] (-.1,-1.1)
  (j2) edge[decorate,decoration={snake,amplitude=.2mm,segment length=1mm,pre length=3pt, post length=1pt}] (.7,-1.1)
 (j2) edge[decorate,decoration={snake,amplitude=.2mm,segment length=1mm,pre length=3pt, post length=1pt}] (.1,-1.1);

\node[fill = teal!60!white, opacity = 0.8, draw, rounded corners=3pt] at (3,2) {$\Phi_i^\circ$};
\node[fill = blue!60!white, opacity = 0.8, draw, rounded corners=3pt] at (3,1) {$\Phi_i^\uparrow$};
\node[fill = red!60!white, opacity = 0.8, draw, rounded corners=3pt] at (3,0) {$\Phi_i^\downarrow$};

\end{tikzpicture} 
\end{subfigure}\quad
\begin{subfigure}[b]{0.55\textwidth}
\begin{align*}
\Phi_i^\circ(\tpl x_i) & = \Big|\sigma_{\tpl X_i = \tpl x_i}\Big(\join\limits_{\substack{j \in [r] \\ j \neq i}} \mat S_j\Big)\Big| \\
\Phi_i^\uparrow(\tpl x_p) & = \Big|\sigma_{\tpl X_p = \tpl x_p}\Big(\join\limits_{\substack{j \in [r] \\ \mat S_j \text{ is not in the subtree of } \mat S_i}} \mat S_j\Big)\Big| \\
\Phi_i^\downarrow(\tpl x_p) & = \Big|\sigma_{\tpl X_p = \tpl x_p}\Big(\join\limits_{\substack{j \in [r] \\ \mat S_j \text{ is in the subtree of } \mat S_i}} \mat S_j\Big)\Big|
\end{align*}
\end{subfigure}
\caption{Count queries used by \Figaro, defined over the join tree $\tau$ of the $r$ input matrices $\mat S_1, \ldots, \mat S_r$. $\tpl X_i$ are the join attributes of $\mat S_i$ and $\tpl X_p$ are the join attributes that $\mat S_i$ has in common with its parent in $\tau$. $\Phi_i^\circ$ gives the number of tuples in the join of all relations except $\mat S_i$, grouped by the attributes $\tpl X_i$. $\Phi_i^\uparrow$ and $\Phi_i^\downarrow$ give the number of tuples in the join of the relations that are in the subtree of $\mat S_i$ in $\tau$ respectively not in this subtree, grouped by the attributes $\tpl X_p$.   \label{fig:counts}}
\end{figure}

We define the count queries $\Phi_i^\circ$, $\Phi_i^\uparrow$ and $\Phi_i^\downarrow$ for each input matrix $\mat S_i$ in a given join tree $\tau$ ($i \in [r]$). Fig.\@~\ref{fig:counts} depicts graphically the meaning of these count queries: They give the sizes of joins of subsets of the input matrices, grouped by various join attributes of $\mat S_i$: $\Phi_i^\uparrow$ and $\Phi_i^\downarrow$ join all matrices above and respectively below $\mat S_i$ in the join tree $\tau$, while $\Phi_i^\circ$ joins all matrices except $\mat S_i$.

Each of these aggregates can be computed individually in time linear in the size of the input matrices using standard techniques for group-by aggregates over acyclic joins~\cite{Larson,Olteanu:FDB:2016}.
This section shows how to further speed up their evaluation by identifying and executing common computation across all these aggregates. This reduces the necessary number of scans of each of the input matrices from $\bigO(r)$ scans to two scans.

We recall our notation: $\tau$ be the join tree; $\tpl X_i$ are the join attributes of $\mat S_i$ and $\tpl X_{ij}$ are the common join attributes of $\mat S_i$ and its child $\mat S_j$ in $\tau$. We assume that: each matrix $\mat S_i$ is sorted on the attributes $\tpl X_i$; and a child $\mat S_j$ of $\mat S_i$ in $\tau$ is first sorted on the attributes $\tpl X_{ij}$.

We first initialize the counts $\FuncSty{rows\_per\_key}_i(\tpl x_i)$ for each matrix $\mat S_i$ and value $\tpl x_i$ for its join attributes $\tpl X_i$. These counts give the number of those rows in $\mat S_i$ that contain the join key $\tpl x_i$ for the columns $\tpl X_i$. This step is done in one pass over the sorted matrices and takes linear time in their size. The aggregates $\Phi_i^\downarrow$ are computed in the same data pass. The computation of the aggregates $\Phi_i^\circ$ and $\Phi_i^\uparrow$ is done in a second pass over the matrices. We utilize the following relationships between the aggregates.

{\bf\noindent Computing $\Phi_i^\downarrow$:} This aggregate gives the size of the join of the matrices in the subtree of $\mat S_i$ in $\tau$ grouped by the join attributes $\tpl X_{p}$ that are common to $\mat S_i$ and its parent in $\tau$.
If $\mat S_i$ is a leaf in $\tau$, then $\tpl X_{p} = \tpl X_i$ and $\Phi_i^\downarrow(\tpl x_i) = \FuncSty{rows\_per\_key}_i(\tpl x_i)$.
If $\mat S_i$ is not a leaf in $\tau$, then we compute the intermediate aggregate
 \[\Theta^\downarrow_{i}(\tpl x_i) =  \Big|\sigma_{\tpl X_i = \tpl x_i}\Big(\join\limits_{\substack{j \in [r] \\ \mat S_j \text{ is in the subtree of } \mat S_i}} \mat S_j\Big)\Big| \]
which gives the size of the join of the same matrices as $\Phi_i^\downarrow$, but grouped by $\tpl X_i$ instead of $\tpl X_{p}$. The reason for computing $\Theta_i^\downarrow$ is that it is useful for other aggregates as well, as discussed later. This aggregate can also be expressed as \[\Theta_i^\downarrow(\tpl x_i) = \FuncSty{rows\_per\_key}_i(\tpl x_i) \cdot \prod_{\text{child }\mat S_j \text{ of } \mat S_i} \Phi_j^\downarrow(\tpl x_{ij}),\] where $\tpl x_{ij}$ is the projection of $\tpl x_i$ to the join attributes $\tpl X_{ij}$.
The aggregate $\Phi_i^\downarrow(\tpl x_{p})$ is obtained by summing all values $\Theta_i^\downarrow(\tpl x_i)$ such that $\tpl x_i$ projected to $\tpl X_p$ equals $\tpl x_p$.

All aggregates $\Phi_1^\downarrow,\ldots,\Phi_r^\downarrow$ can thus be computed in a bottom-up traversal of the join tree $\tau$ and in one pass over the sorted matrices.

{\bf\noindent Computing  $\Phi_i^\uparrow$:} This aggregate is defined similarly to $\Phi_i^\downarrow$, where the join is now over all matrices that are \emph{not} in the subtree of $\mat S_i$ in the join tree $\tau$. It is computed in a top-down traversal of $\tau$.

Assume $\Phi_i^\uparrow(\tpl x_{p})$ is already computed, where $\tpl x_p$ is a value for the join attributes $\tpl X_p$ that $\mat S_i$ has in common with its parent in $\tau$. The intermediate aggregate $\FuncSty{full\_join\_size}_i(\tpl x_i) = \Phi_i^\uparrow(\tpl x_{p}) \cdot \Theta_i^\downarrow(\tpl x_i)$ gives  the size of the join over all matrices, grouped by $\tpl X_{i}$.
For the root $\mat S_1$ of $\tau$, we set $\FuncSty{full\_join\_size}_1(\tpl x_1) = \Theta_1^\downarrow(\tpl x_1)$.

We next define aggregates $\FuncSty{full\_join\_size}_{ij}(\tpl x_{ij})$ that give the size of the join of all matrices for each value of the join attributes $\tpl X_{ij}$ that are common to $\mat S_i$ and its child $\mat S_j$. They can be computed by summing up the values $\FuncSty{full\_join\_size}_i(\tpl x_i)$ for the keys $\tpl x_i$ that agree with $\tpl x_{ij}$ on $\tpl X_{ij}$.
Since this size is the product of $\Phi_j^\downarrow(\tpl x_{ij})$ and $\Phi_j^\uparrow(\tpl x_{ij})$, we have $\Phi_j^\uparrow(\tpl x_{ij}) =  \FuncSty{full\_join\_size}_{ij}(\tpl x_{ij})/\Phi_j^\downarrow(\tpl x_{ij})$, where $\Phi_j^\downarrow$ is already computed.
For all children $\mat S_j$ of $\mat S_i$, we can compute all sums $\FuncSty{full\_join\_size}_{ij}$ together in one pass over $\mat S_i$.

\begin{algorithm}[!tp]
  \DontPrintSemicolon%
  \FuncSty{compute-counts}(matrices $\mat S_1, \ldots, \mat S_r$, join tree $\tau$)
\Begin{
\Comment{$\Phi_i^\downarrow, \Phi_i^\uparrow, \Phi_i^\circ, \FuncSty{rows\_per\_key}_i$ and $\Theta_i^\downarrow$ are initially empty ordered maps, for every matrix $\mat S_i$ (for the root $\mat S_r$ of $\tau$, $\Phi_r^\downarrow, \Phi_r^\uparrow$ do not exist).}
    \FuncSty{pass$_1$}($1$) \\
    \FuncSty{pass$_2$}($1$)
}

\FuncSty{pass$_1$}\,(index $i$)
\Begin{
\Comment{Computes $\Phi_i^\downarrow$ and intermediate maps $\FuncSty{rows\_per\_key}, \Theta_i^\downarrow$.}
\ForEach{child $\mat S_j$ of $\mat S_i$ in $\tau$} {
 \FuncSty{pass$_1$}($j$)
}
 \ForEach{value $\tpl x_i$ of the join attributes $\tpl X_i$ of $\mat S_i$} {\label{alg4:loop1}
  $\FuncSty{rows\_per\_key}_i(\tpl x_i) \gets |\sigma_{\tpl X_i = \tpl x_i} \mat S_i|$ \\
  $\Theta_i^\downarrow(\tpl x_i) \gets \FuncSty{rows\_per\_key}_i(\tpl x_i)$ \\
  \ForEach{child $\mat S_j$ of $\mat S_i$ in $\tau$} {
  \Comment{let $\tpl x_{ij}$ be the projection of $\tpl x_i$ onto $\tpl X_{ij}$, the join attributes shared between $\mat S_i$ and $\mat S_j$}
   $\Theta_i^\downarrow(\tpl x_i) \gets \Theta_i^\downarrow(\tpl x_i) \cdot \Phi_j^\downarrow(\tpl x_{ij})$
  }
  \If{$\mat S_i$ is not the root of $\tau$}{
  \Comment{let $\tpl x_{p}$ be the projection of $\tpl x_i$ onto the join attributes shared between $\mat S_i$ and its parent in $\tau$}
  $\Phi_i^\downarrow(\tpl x_p) \gets \Phi_i^\downarrow(\tpl x_p) + \Theta_i^\downarrow(\tpl x_i)$\label{alg4:write1}
  }
 }
}

\FuncSty{pass$_2$}\,(index $i$)
\Begin{
\Comment{Computes $\Phi_i^\uparrow$ and $\Phi_i^\circ$.}
 \ForEach{value $\tpl x_i$ of the join attributes $\tpl X_i$ of $\mat S_i$} {\label{alg4:loop2}
 \eIf{$\mat S_i$ is not the root of $\tau$}{
 \Comment{let $\tpl x_{p}$ be the projection of $\tpl x_i$ onto the join attributes shared by $\mat S_i$ and its parent in $\tau$}
  $\DataSty{up\_count} \gets \Phi_i^\uparrow(\tpl x_{p})$
 }{
 $\DataSty{up\_count} \gets 1$
 }
   $\FuncSty{full\_join\_size}_i(\tpl x_i) \gets \Theta_i^\downarrow(\tpl x_i) \cdot \DataSty{up\_count}$ \\
   \ForEach{child $\mat S_j$ of $\mat S_i$ in $\tau$} {
   \Comment{let $\tpl x_{ij}$ be the projection of $\tpl x_i$ onto $\tpl X_{ij}$}
    $\Phi_j^\uparrow(\tpl x_{ij}) \gets \Phi_j^\uparrow(\tpl x_{ij}) + \FuncSty{full\_join\_size}_i(\tpl x_i)$\label{alg4:write2}
   }
    $\Phi_i^\circ(\tpl x_i) \gets \frac{\FuncSty{full\_join\_size}_i(\tpl x_i)}{\FuncSty{rows\_per\_key}_i(\tpl x_i)}$
 }
 \ForEach{child $\mat S_j$ of $\mat S_i$ in $\tau$} {
  \ForEach{value $\tpl x_{ij}$ of join attributes $\tpl X_{ij}$ of $\mat S_j$} {
   $\Phi_j^\uparrow(\tpl x_{ij}) \gets \frac{\Phi_j^\uparrow(\tpl x_{ij})}{\Phi_j^\downarrow(\tpl x_{ij})}$
  }
  \FuncSty{pass$_2$}($j$)
 }
}
  \caption{Computing the count aggregates for \Figaro.}
  \label{alg:counts}
\end{algorithm}

{\bf\noindent Computing $\Phi_i^\circ$:} This aggregate gives the size of the join of all matrices except $\mat S_i$, grouped by the attributes $\tpl X_i$.
If $\mat S_i$ is a leaf in $\tau$, then $\Phi_i^\circ = \Phi_i^\uparrow$ by definition.
For a non-leaf $\mat S_i$, we re-use the values $\FuncSty{full\_join\_size}_i(\tpl x_i)$ defined above. This value is very similar to $\Phi_i^\circ(\tpl x_i)$, but also depends on the size $\FuncSty{rows\_per\_key}_i(\tpl x_i)$. More precisely, we need to divide $\FuncSty{full\_join\_size}_i(\tpl x_i)$ by $\FuncSty{rows\_per\_key}_i(\tpl x_i)$ to obtain $\Phi_i^\circ(\tpl x_i)$.

Alg.\@~\ref{alg:counts} gives the procedure \FuncSty{compute-counts} for shared computation of the three aggregate types. First, in a bottom-up pass of the join tree $\tau$, it computes the aggregates $\Theta_i^\downarrow$ and $\Phi_i^\downarrow$. Then, it computes $\Phi_i^\uparrow$ and $\Phi_i^\circ$ in a top-down pass over $\tau$.
The view $\FuncSty{full\_join\_size}_i$ is defined depending on whether $\mat S_i$ is the root. For every child $\mat S_j$, the value $\FuncSty{full\_join\_size}_i$ is added to $\Phi_j^\uparrow(\tpl x_{ij})$. The division by $\Phi_j^\downarrow(\tpl x_{ij})$ is done just before the recursive call. Before that call, $\Phi_i^\circ$ is computed.

\begin{lemma}\label{lem:counts}
Given the matrices $\mat S_1, \ldots, \mat S_r$ and the join tree $\tau$, the aggregates $\Phi_i^\downarrow$, $\Phi_i^\uparrow$ and $\Phi_i^\circ$ from Fig.\@~\ref{fig:counts} can be computed in time $\bigO(|\mat S_1|+\cdots+|\mat S_r|)$ and with two passes over these matrices.
\end{lemma}

We parallelize Alg.\@~\ref{alg:counts} by executing the loops starting in lines~\ref{alg4:loop1} and~\ref{alg4:loop2}  in parallel for different values $\tpl x_i$. Atomic operations are used for the assignments in lines~\ref{alg4:write1} and~\ref{alg4:write2} to handle concurrent write operations of different threads.

\section{\texorpdfstring{\Figaro: Pushing rotations past joins}{Figaro: Pushing rotations past joins}}\label{sec:figaro}

We are now ready to introduce \Figaro. Alg.\@~\ref{alg:figaro} gives its pseudocode. It takes as input the matrices $\mat S_1, \ldots, \mat S_r$ and a join tree $\tau$ of the natural join of the input matrices. It computes an almost upper-triangular matrix $\mat R_0$ with at most $|\mat S_1| + \cdots + |\mat S_r|$ non-zero rows such that $\mat R_0$ can be alternatively obtained from the natural join of the input matrices by a sequence of Givens rotations.
It does so in linear time in the size of the input matrices.
In contrast to the example in Sec.\@~\ref{sec:naive}, \Figaro does not require the join output $\mat A$ to be materialized. Instead, it traverses the input join tree bottom-up and computes heads and tails of the join of the matrices under each node in the join tree.

\Figaro uses the following two high-level ideas.

First, the join output typically contains multiple copies of an input row, where the precise number of copies depends on the number of rows from other matrices that share the same join key. When using Givens rotations to set all but the first occurrence of a value to zero, this first occurrence will be scaled by some factor, which is given by a count query (Sec.\@~\ref{sec:counts}). \Figaro applies the scaling factor directly to the value, without the detour of constructing the multiple copies and getting rid of them again.

Second, all matrix head and tail computations can be done independently on the different columns of the input matrix. \Figaro performs them on the input matrix, where these columns originate, and combines the results, including the application of scaling factors, in a bottom-up pass over the join tree $\tau$.

\begin{algorithm}[!tp]
  \DontPrintSemicolon

\Figaro(input matrices $\mat S_1, \ldots, \mat S_r$, join tree $\tau$, index $i$)
  \Begin{
  \Comment{$\Result$, $\Data$ are matrices, $\XScales$ is a vector. All are initially empty.}
    \FuncSty{heads\_and\_tails}\;
    \If{$\mat S_i$ is not a leaf in $\tau$}{
      \FuncSty{process\_and\_join\_children}\;
      \If{$\mat S_i$ is not the root of $\tau$}{\label{alg2label:notroot}
      \FuncSty{project\_away\_join\_attributes}}
    }
     \If{$\mat S_i$ is the root of $\tau$}{$\Result.\FuncSty{append}(\Data)$ \label{alg2label:res3}\;
\Output $\Result$}
    \Output $(\Result, \Data, \XScales)$
    }

\FuncSty{heads\_and\_tails}:
\Begin{
\Comment{Compute heads and tails of $\mat S_i$, grouped by the columns $\tpl X_i$. Heads are written to $\Data$, tails are scaled and written to $\Result$.}
 \ForEach{value $\tpl x_i$ of the join attributes $\tpl X_i$ in $\mat S_i$}{\label{alg2label:loop1}
     \Comment{Let $\mat S^{\tpl x_i}_i$ consist of all rows of $\mat S_i$ with value $\tpl x_i$ for columns $\tpl X_i$, projected onto the data attributes $\tpl Y_i$}
      \Result.\FuncSty{append}($\begin{bmatrix}
      \mat T_1 & \cdots & \mat T_r
     \end{bmatrix}$)  \label{alg2label:res1} \Begin{
      where $\mat T_k = \mat 0$ for $k \neq i$ and $\mat T_i = \mbox{$\calT(\mat S^{\tpl x_i}_i)$} \cdot \sqrt{\Phi^\circ_i(\tpl x_i)}$
     }
     $\Data[\tpl x_i : \tpl Y_i] \gets \calH(\mat S^{\tpl x_i}_i)$\;
     $\XScales[\tpl x_i] \gets \sqrt{|\mat S^{\tpl x_i}_i|}$
  }
}

\FuncSty{process\_and\_join\_children}:
  \Begin{
  \Comment{Recursively applies \Figaro to all children. Concatenates the results $\Result$, joins the matrices $\Data$ and applies the factors $\XScales$}
      \ForEach{child $\mat S_j$ of $\mat S_i$ in $\tau$}{
       $(\Result_j, \Data_j, \XScales_j) \gets \Figaro(\mat S_1, \ldots, \mat S_r,\tau,j)$ \;
       \Result.\FuncSty{append}($\Result_j$)
      }
      \ForEach{value $\tpl x_i$ of the join attributes $\tpl X_i$ in $\mat S_i$}{\label{alg3label:loop}
       \Comment{let $\tpl X_{ij}$ be the join attributes shared between $\mat S_i$ and $\mat S_j$ and let $\tpl x_{ij}$ be the projection of $\tpl x_i$ onto $\tpl X_{ij}$.}
       \ForEach{child $\mat S_j$ of $\mat S_i$ in $\tau$}{
        \ForEach{$\mat S_k$ in the subtree of $\mat S_j$ in $\tau$}{
          $\Data[\tpl x_i : \tpl Y_k] \gets \mbox{$\Data_j[\tpl x_{ij} : \tpl Y_k]$} \cdot \XScales[\tpl x_{i}] \cdot$ \; \Indp \nonl  $\prod\limits_{j' \in [r] \setminus \{j\}}  \XScales_{j'}[\tpl x_{i{j'}}]$
         }
        }
        $\Data[\tpl x_i : \tpl Y_i] \gets \Data[\tpl x_i : \tpl Y_i] \cdot \prod\limits_{j \in [r]}  \XScales_j[\tpl x_{ij}]$ \;
        $\XScales[\tpl x_i] \gets \XScales[\tpl x_i] \cdot \prod\limits_{j \in [r]}  \XScales_j[\tpl x_{ij}]$
      }
}

\FuncSty{project\_away\_join\_attributes}:
  \Begin{
\Comment{Let $\tpl X_{p}$ be the join attributes shared between $\mat S_i$ and its parent in $\tau$. So far, $\Data$ has one row for every value of $\tpl X_i$. Reduce this to one row for every value of $\tpl X_{p}$.}
     \ForEach{value $\tpl x_p$ of the join attributes $\tpl X_p$ in $\mat S_i$}{\label{alg2label:for}
     \Result.\FuncSty{append}($\begin{bmatrix}
      \mat T_1 & \cdots & \mat T_r
     \end{bmatrix}$) \label{alg2label:res2}   \Begin{
      where $\mat T_k = \mat 0$ if $\mat S_k$ is not in the subtree of $\mat S_i$ in $\tau$; else \;
      $\mat T_k = \mbox{$\calT(\Data[\tpl x_{p}, \ast : \tpl Y_k], \XScales[\tpl x_{p}, \ast])$} \cdot \sqrt{\Phi^\uparrow_i(\tpl x_p)}$
     }
      $\Data'[\tpl x_{p} : ] \gets \calH(\Data[\tpl x_p, \ast : ], \XScales[\tpl x_p, \ast])$ \;
      $\XScales'[\tpl x_{p}] \gets \sqrt{\Phi^\downarrow_i(\tpl x_{p})}$ \;
     }
     $(\Data, \XScales) \gets (\Data', \XScales')$
}

  \caption{\Figaro computes the almost upper-triangular matrix $\mat R_0$ from the input matrices $\mat S_1, \ldots, \mat S_r$ with join tree $\tau$.}
  \label{alg:figaro}
\end{algorithm}

\textbf{The algorithm.}
\Figaro manages two matrices $\Data$ and $\Result$. The matrix $\Result$ holds the result of \Figaro, the matrix $\Data$ holds (generalized) heads that will be processed further in later steps.
The algorithm proceeds recursively along the join tree $\tau$, starting from the root.
It maintains the invariant that after the computation finished for a non-root node $\mat S_i$ of the join tree, $\Data$ contains exactly one row for every value of the join attributes that are shared between $\mat S_i$ and its parent in $\tau$.

For a matrix $\mat S_i$, \Figaro first iterates over all values $\tpl x_i$ of its join attributes $\tpl X_i$. For each $\tpl x_i$ it computes head and tail of the matrix $\mat S_i^{\tpl x_i}$ that is obtained by selecting all rows of $\mat S_i$ that have the join key $\tpl x_i$ and then by projecting these rows onto the data columns $\tpl Y_i$.
The tail is multiplied with the scaling factor $\Phi_i^\circ(\tpl x_i)$ and written to the output \Result, adding zeros to other columns.
The head is stored in the matrix $\Data$. The scaling factor $\sqrt{|\mat S_i^{\tpl x_i}|}$ that, as per Lemma \ref{lem:cartesianproduct}, is applied to other columns, is stored in the vector $\XScales$.
If $\mat S_i$ is a leaf in the join tree $\tau$, then the invariant is satisfied.

If $\mat S_i$ is not a leaf in the join tree $\tau$, then the algorithm is recursively called for its children. Any output of the recursive calls is written to \Result.
The recursive call for the child $\mat S_j$ also returns a matrix $\Data_j$ that contains columns $\tpl Y_k$ with computed heads corresponding to this column, for all $k$ such that $\mat S_k$ is in the subtree of $\mat S_j$ in $\tau$, as well as scaling factors $\XScales_j$. As to the maintained invariant, $\Data_j$ contains exactly one row for every value of the join attributes $\tpl X_{ij}$ that are shared between $\mat S_i$ and $\mat S_j$.

The contents of the matrices $\Data$ and $\Data_j$ are joined.
More precisely, for each join value $\tpl x_i$ of the attributes $\tpl X_i$ of $\mat S_i$, the entry $\Data[\tpl x_i : \tpl Y_k]$ is set to the entry $\Data_j[\tpl x_{ij} : \tpl Y_k]$, where $\tpl x_{ij}$ is the projection of $\tpl x_i$ onto $\tpl X_{ij}$.
Also, scaling factors are applied to the different columns of $\Data$. For the columns $\tpl Y_k$, all scaling factors from $\XScales$ and from the vectors $\XScales_j$ need to be applied, for all $j$ such that $\mat S_k$ is not in the subtree of $\mat S_j$. In the case of $\tpl Y_i$, all factors from the vectors $\XScales_j$ are applied, but not from $\XScales$.

If $\mat S_i$ is not the root of the join tree, rows of $\Data$ are aggregated in the procedure \FuncSty{project\_away\_join\_attributes} to satisfy the invariant.
Before entering the loop of Line~\ref{alg2label:for}, $\Data$ contains one row for every value $\tpl x_i$ of the join attributes $\tpl X_i$ of $\tpl S_i$. The number of rows is reduced in \FuncSty{project\_away\_join\_attributes} such that there is only one row for every value $\tpl x_p$ of the join attributes $\tpl X_p$ that are shared between $\mat S_i$ and its parent in $\tau$.
To do so, \Figaro iterates over all values $\tpl x_p$ of these attributes. For each $\tpl x_p$ it takes the rows of $\Data$ for the keys $\tpl x_i$ that agree with $\tpl x_p$ on $\tpl X_p$, and computes their generalized head and tail.
The generalized tail is scaled by the factor $\sqrt{\Phi_i^\uparrow(\tpl x_p)}$ and written to $\Result$, adding zeros for other columns.
The matrix \Data is overwritten with the collected generalized heads. So, the invariant is satisfied now.
The scaling factor $\sqrt{\Phi_i^\downarrow(\tpl x_p)}$ equals the scaling factor that is applied to other columns as per Lemma~\ref{lem:scaledcartesianproduct}, and is stored.

The next theorem states that indeed the result of \Figaro is an almost upper-triangular matrix with a linear number of non-zero rows. It also states the runtime guarantees of \Figaro.

\begin{theorem}\label{thm:figaro}
Let the matrices $\mat S_1, \ldots, \mat S_r$ represent fully reduced relations and let $\tau$ be any join tree of the natural join over these relations. Let $\mat A$ be the matrix representing this natural join. Let $M$ be the overall number of rows and $N$ be the overall number of data columns in the input matrices $\mat S_i$.

On input $(\mat S_1, \ldots, \mat S_r$,$\tau$,$1$), \Figaro returns a matrix $\mat R_0$ in $\bigO(MN)$ time such that:
\begin{enumerate}[(1)]
 \item $\mat R_0$ has at most $M$ rows,
 \item there is an orthogonal matrix $\mat Q$ such that \[\mat A[: \tpl Y] = \mat Q \begin{bmatrix}
 \mat R_0 \\
 \mat 0
 \end{bmatrix}.\]
\end{enumerate}

\end{theorem}

We can parallelize \Figaro by executing the loop starting in line~\ref{alg2label:loop1} of Alg.\@~\ref{alg:figaro} in parallel for different values $\tpl x_i$, similarly for the loops starting in lines~\ref{alg2label:for} and~\ref{alg3label:loop}. As the executions are independent, no synchronization between the threads is necessary.

\textbf{Comparison with Sec.\@~\ref{sec:naive}.}
We exemplify the execution of \Figaro using the same input as in Sec.\@~\ref{sec:naive} and depicted in Fig.\@~\ref{fig:examplenaive}. We will obtain the same result $\mat R_0$ as in that section, modulo permutations of rows.

After the initial call of $\Figaro(\mat S_1, \ldots, \mat S_4,\tau,1)$, at first heads and tails of $\mat S_1$ are computed. We skip the explanation of this part and focus on the following recursive call of $\Figaro(\mat S_1, \ldots, \mat S_4,\tau,2)$.
When computing the heads and tails of $\mat S_2$, at first the matrices $\sqrt{\Phi^\circ_2(a_jb_kc_\ell)}\calT(\mat S_2^{a_jb_kc_\ell})$ are padded with zeros and written to $\Result$, for all $j, k, \ell \in \{1,2\}$.
Then, the entry $\Data[a_jb_kc_\ell : Y_3]$ is set to $\calH(\mat S_2^{a_jb_kc_\ell})$, $\XScales[a_jb_kc_\ell]$ becomes $\sqrt{|\mat S_2^{a_jb_kc_\ell}|}$.

Next, the children $\mat S_3$ and $\mat S_4$ of $\mat S_2$ in $\tau$ are processed and the intermediate results are joined. The recursive calls for the children return  $\Data_3[b_k : Y_3] = \calH(\mat S_3^{b_k})$ and $\XScales_3[b_k] = \sqrt{|\mat S_3^{b_k}|}$ as well as $\Data_4[c_\ell : Y_4] = \calH(\mat S_4^{c_\ell})$ and $\XScales_4[c_\ell] = \sqrt{|\mat S_4^{c_\ell}|}$; the matrices $\Result_3$, $\Result_4$ contain as non-zero entries $\sqrt{\Phi^\circ_3(b_k)} \calT(\mat S_3^{b_k})$ and $\sqrt{\Phi^\circ_4(c_\ell)} \calT(\mat S_4^{c_\ell})$, respectively, for every $k, \ell \in \{1,2\}$.

After joining the results and applying the scaling factors, the rows \mbox{$\Data[a_jb_kc_\ell :]$} are as follows for every $j, k, \ell \in \{1,2\}$, with $\beta_{jk \ell} = \sqrt{|\mat S_2^{a_jb_kc_\ell}|\cdot|\mat S_3^{b_k}|\cdot|\mat S_4^{c_\ell}|}$:

\begin{center}
\begin{tabular}{ccccc}
$Y_2$ & $Y_3$ & $Y_4$  \\
\hline
\small $\frac{\beta_{j k \ell}}{\sqrt{|\mat S_2^{a_j b_k c_\ell}|}} \calH(\mat S_2^{a_j b_k c_\ell}) $  &  \small $\frac{\beta_{j k \ell}}{\sqrt{|\mat S_3^{b_k}|}} \calH(\mat S_3^{b_k})$  &  \small $\frac{\beta_{j k \ell}}{\sqrt{|\mat S_4^{c_\ell}|}} \calH(\mat S_4^{c_\ell})$     
\end{tabular}
\end{center}
It holds $\XScales[a_j b_kc_\ell] =\beta_{j k \ell}$.

Notice the similarity to the corresponding rows from Sec.\@~\ref{sec:naive}. There we obtained, for $j=1$ and with $\alpha_{k \ell} = \sqrt{|\mat S_1^{a_1}|\cdot|\mat S_2^{a_1 b_k c_\ell}|\cdot|\mat S_3^{b_k}|\cdot|\mat S_4^{c_\ell}|}$, rows that have a column $Y_1$ consisting of $\frac{\alpha_{k \ell}}{\sqrt{|\mat S_1^{a_1}|}} \calH(\mat S_1^{a_1})$, and the columns

\begin{center}
\setlength{\tabcolsep}{5pt}
\begin{tabular}{ccccc}
$Y_2$ & $Y_3$ & $Y_4$  \\
\hline
$\frac{\alpha_{k \ell}}{\sqrt{|\mat S_2^{a_1 b_k c_\ell}|}} \calH(\mat S_2^{a_1 b_k c_\ell}) $  &  $\frac{\alpha_{k \ell}}{\sqrt{|\mat S_3^{b_k}|}} \calH(\mat S_3^{b_k})$  &  $\frac{\alpha_{k \ell}}{\sqrt{|\mat S_4^{c_\ell}|}} \calH(\mat S_4^{c_\ell}).$    
\end{tabular}
\end{center}

For $j = 1$, the only difference in the columns $Y_2$ to $Y_4$ is in the scaling factors $\alpha_{k \ell}$ and $\beta_{1 k \ell}$, which only differ by the factor $\sqrt{|\mat S_1^{a_1}|}$.

The join attributes $X_{23}$ and $X_{24}$ are then projected away, as they do not appear in the parent $\mat S_1$ of $\mat S_2$.
Observe that the vectors that are used for defining the generalized heads and tails are the same here and in Sec.\@~\ref{sec:naive}. For example, the part of $\XScales$ that corresponds to $a_j = a_1$ equals $\mat v = \begin{bmatrix}
 \sqrt{|\mat S_2^{a_1b_1c_1}|\cdot|\mat S_3^{b_1}|\cdot|\mat S_4^{c_1}|} \\
 \vdots \\
 \sqrt{|\mat S_2^{a_1b_2c_2}|\cdot|\mat S_3^{b_2}|\cdot|\mat S_4^{c_2}|}
\end{bmatrix}$, the vector used in Sec.\@~\ref{sec:naive}.
So, the generalized heads and tails for the rows of $\Data$ associated with join keys $a_1$ and $a_2$, respectively, have the same results as in Sec.\@~\ref{sec:naive} for the columns $Y_2$ to $Y_4$, modulo the factor $\sqrt{|\mat S_1^{a_1}|}$. 
This factor equals $\sqrt{\Phi_2^\uparrow(a_1)}$, which \Figaro applies to the generalized tails. 
That factor is also applied to the generalized heads in $\Data$, namely by the procedure \FuncSty{process\_and\_join\_children} in the call $\Figaro(\mat S_1, \ldots, \mat S_4,\tau,1)$, after the call of \Figaro for $\mat S_2$ terminates. There, also the column $Y_1$ is added.
The rows in the final result of $\Figaro(\mat S_1, \ldots, \mat S_4,\tau,1)$ are then the same as in Sec.\@~\ref{sec:naive}.

\section{Post-processing}
\label{sec:postprocessing}

The output of \Figaro is so far not an upper-triangular matrix, but a matrix $\mat R_0$ consisting of linearly many rows in the size of the input matrices.
The transformation of $\mat R_0$ into the final upper-triangular matrix $\mat R$ is the task of a post-processing step that we describe in this section.
This step can be achieved by general techniques for QR factorization, such as Householder transformations and textbook Givens rotations approaches.
The approach we present here exploits the structure of $\mat R_0$, which has large blocks of zeros.

The non-zero blocks of $\mat R_0$ are either tails of matrices (see line~\ref{alg2label:res1} of Alg.\@~\ref{alg:figaro}), generalized tails of sets of matrices (line~\ref{alg2label:res2}), or the content of the matrix $\Data$ at the end of the execution of \Figaro (line~\ref{alg2label:res3}).
In a first post-processing step, these blocks are upper-triangularized individually and potentially arising all-zero rows are discarded. %
In a second post-processing step, the resulting block of rows is transformed into the upper-triangular matrix $\mat R$.

In both steps we need to make blocks of rows upper-triangular, i.e., we need to compute their QR decomposition. This can be done using off-the-shelf Householder transformations. We implemented 
an alternative method dubbed THIN \cite{golub1986parallel,da2002new}. It first divides the rows of a block among the available threads, each thread then applies a sequence of Givens rotations to bring its share of rows into upper-triangular form. Then THIN uses a classical parallel Givens rotations approach \cite[p.~257]{matrix2016comp} on the collected intermediate results to obtain the final upper-triangular matrix.

All approaches need time $\bigO(MN^2)$, where $M$ is the number of rows and $N$ is the number of columns of the input $\mat R_0$ to post-processing. While the main part of \Figaro works in linear time in the overall number of rows and columns of the input matrices (Theorem~\ref{thm:figaro}), post-processing is only linear in the number of rows.

\section{\texorpdfstring{From $\mat R$ to $\mat Q$, SVD and PCA}{From R to Q, SVD and PCA}}
\label{sec:beyondR}

The previous sections focus on computing the upper-triangular matrix $\mat R\in\mathbb{R}^{n\times n}$ in the QR decomposition of the matrix $\mat A\in\mathbb{R}^{m\times n}$ representing the join output with $m$ tuples and $n$ data attributes. Using $\mat R$ and the \emph{non-materialized} join matrix $\mat A$, this section shows how to compute the orthogonal matrix $\mat Q$ in the QR decomposition of $\mat A$, the singular value decomposition of $\mat A$, and the principal component analysis of $\mat A$.

\subsection{\texorpdfstring{The Orthogonal Matrix $\mat Q$}{The Orthogonal Matrix Q}}
\label{sec:q}

Using $\mat A = \mat Q \mat R$, we can compute the orthogonal matrix $\mat Q\in\mathbb{R}^{m\times n}$ as follows: $\mat Q = \mat A \mat R^{-1}$. The upper triangular matrix $\mat R$ admits an inverse, which is also upper triangular, in case the values along the diagonal are non-zero. To compute $\mat Q$, we do not need to materialize $\mat A$! Each row in $\mat A$ represents one tuple in the join result. We can enumerate the rows in $\mat A$ without materializing $\mat A$ using factorization techniques from prior work~\cite{Olteanu:FDB:2015,Olteanu:FDB:2016}. This enumeration has the delay constant in the size of the input database and linear in the number $n$ of the data attributes. The delay is the maximum of three times: the time to output the first tuple, the time between outputting one tuple and outputting the next tuple, and the time to finish the enumeration after the last tuple was outputted. Before the enumeration starts, we calibrate the input relations by removing the dangling tuples, i.e., tuples that do not contribute to any output tuple. For acyclic joins (as considered in this article), this calibration takes  time linear in the size of the input database~\cite{AbiteboulHV95}.
To enumerate, we keep an iterator over the tuples of each relation. We construct an output tuple in a top-down traversal of the join tree, where for each tuple at the iterator of a relation $P$ we consider the possible matching tuples at the iterators of the relations that are children of $P$ in the join tree.

Let $\mat R^{-1} = [\mat r_1 \cdots \mat r_n]$, where $\mat r_1, \ldots, \mat r_n$ are column vectors. Let the rows in $\mat A$ be $\mat t_1,\ldots, \mat t_m$. The cell $\mat Q(i,j)$ is the dot product  $\langle \mat t_i,\mat r_j\rangle$. Once $\mat R^{-1}$ is computed in $O(n^3)$ time, we can enumerate the cells in $\mat Q$ with delay constant in the size of the input database and linear in the number $n$ of data attributes. To construct the entire matrix $\mat Q$, it then takes $O(mn^2)$ time.

We further use an optimization that pu\-shes the computation of $\mat Q$ past the enumeration of the join output to reduce the complexity from $O(mn^2)$ to $O(mn\ell)$, where $\ell$ is the number of relations in the database. The idea is to compute dot products of each tuple in each relation with selected fragments of each column vector in $\mat R^{-1}$. Such dot products can then be reused in the computation of many cells in $\mat Q$, as explained next.

We assign a unique index $i\in[n]$ to each of the $n$ data attributes. Let $I$ be the set of indices of the data attributes in an input relation. For each tuple $\mat s$ in that relation, we compute the dot products $\langle \mat s(I), \mat r_1(I)\rangle,\ldots,$ $\langle \mat s(I), \mat r_n(I)\rangle$, where $\mat s(I)$ is the vector of the values for data attributes in $\mat s$ and $\mat r_i(I)$ is the vector of those values in $\mat r_i$ at indices in $I$. Let us denote these dot products computed using $\mat s$ as $\mat s.d_1,\ldots,\mat s.d_n$. The cell $(i,j)$ in $\mat Q$ is then the sum of the $j$-th
dot products for the input tuples $\mat s_1,\ldots,\mat s_\ell$ that make up the $i$-th output tuple in the enumeration: $\mat Q(i,j) = \sum_{k\in[\ell]}\mat s_k.d_j$.

In case of categorical data attributes, such dot products can be performed even faster. Assume a data attribute $X$ with $k$ distinct categories, and a tuple $\mat s$ that has at position $i$ the $\ell$-th value of $X$. If we were to one-hot encode $X$ in $\mat s$, then $\mat s$ would include a vector $\mat v$ of $k$ values in place of this category, where $\mat v(\ell)=1$ and $\mat v(\ell')=0$ for $\ell'\in[k], \ell'\neq\ell$. Given a $k$-dimensional vector $\mat r$, which is part of a vector in $\mat R^{-1}$, the dot product $\langle \mat v, \mat r \rangle$ is then $\mat r(\ell)$. This can be achieved even without an explicit one-hot encoding of $X$, the index $\ell$ suffices.

\subsection{Singular Value Decomposition}
\label{sec:svd}

The singular value decomposition of $\mat A\in\mathbb{R}^{m\times n}$ is given by $\mat A = \mat U \mat \Sigma \mat V^\transpose$, where $\mat U\in\mathbb{R}^{m\times n}$ is the orthogonal matrix of left singular vectors, $\mat \Sigma\in\mathbb{R}^{n\times n}$ is the diagonal matrix with the singular values along the diagonal, and $\mat V\in\mathbb{R}^{n\times n}$ is the orthogonal matrix of right singular vectors.
We can compute the SVD of $\mat A$ without materializing $\mat A$ using \Figaro as follows.

The first step computes the upper triangular matrix $\mat R\in\mathbb{R}^{n\times n}$ in the QR decomposition of $\mat A$: $\mat A = \mat Q \mat R$.

The second step computes the SVD of $\mat R$ as: $\mat R = \mat U_R \mat \Sigma_R \mat V_R^\transpose$, where $\mat U_R, \mat \Sigma_R, \mat V_R\in\mathbb{R}^{n\times n}$.
There are several off-the-shelf SVD algorithms~\cite{cline2006computation}, we used the divi\-de-and-conquer algorithm~\cite{gu1995divide} as this was numerically the most accurate in our experiments. Note that computing the SVD takes time $O(n^3)$ for $\mat R\in\mathbb{R}^{n\times n}$ and $O(mn^2)$ for $\mat A\in\mathbb{R}^{m\times n}$. The former computation time is much less than the latter, since $n$ is the number of data columns in the database whereas $m \gg n$ is the size of the join output.

The SVD of $\mat A$ can then be computed using the SVD of $\mat R$ as follows: $\mat A  = \mat U \mat \Sigma \mat V^\transpose = \mat Q \mat R = \mat Q \mat U_R \mat \Sigma_R \mat V_R^\transpose$, where $\mat U = \mat Q \mat U_R$ is orthogonal as it is the multiplication of two orthogonal matrices, $\mat \Sigma = \mat \Sigma_R$, and $\mat V = \mat V_R$.
 This means that the singular values of $\mat R$, as given by $\mat \Sigma_R$, are also the singular values of $\mat A$. The right singular vectors of $\mat A$ are also those of $\mat R$, as given by $\mat V_R$.

 The third and final step computes the left singular vectors of $\mat A$:
\begin{align*}
    \mat U &= \mat Q \mat U_R = \mat A \mat R^{-1} \mat U_R = \mat A (\mat U_R \mat \Sigma_R \mat V_R^\transpose)^{-1} \mat U_R \\
    &= \mat A \mat V_R \mat \Sigma_R^{-1} \mat U_R^{-1} \mat U_R = \mat A \mat V_R \mat \Sigma_R^{-1},
\end{align*}
where we use that $\mat U_R^{-1} \mat U_R = \mat I_n$. The matrices $\mat V_R$ and $\mat\Sigma_R$ are as computed in the previous step. The inverse $\mat\Sigma_R^{-1}$ of the diagonal matrix $\mat\Sigma_R$ is a diagonal matrix that has the diagonal values $\sigma_{i,i}^{-1}$ corresponding to the diagonal values $\sigma_{i,i}$ in $\mat\Sigma_R$. The multiplication $\mat S = \mat\Sigma_R^{-1} \mat V_R$ takes time $O(n^2)$. We are then left to perform the multiplication of the non-materialized matrix $\mat A$ with the $n\times n$ matrix $\mat S$, for which we proceed as for computing $\mat Q$ in Sec.\@~\ref{sec:q}.

\subsection{Principal Component Analysis}

Following standard derivations, we can compute the principal components (PCs) of $\mat A$ using the computation of the SVD of $\mat A$ explained in Sec.\@~\ref{sec:svd}.

We compute the eigenvalues and eigenvectors of
\[\mat A^\transpose \mat A = (\mat U \mat \Sigma \mat V^\transpose)^\transpose \mat U \mat \Sigma \mat V^\transpose = \mat V \mat \Sigma \mat U^\transpose \mat U \mat \Sigma \mat V^\transpose = \mat V \mat \Sigma^2 \mat V^\transpose\]
The sought eigenvalues and their corresponding eigenvectors are the squares of the singular values and respectively the right singular vectors of $\mat A$, or equivalently of the upper-triangular matrix $\mat R$ computed by \Figaro. The PCs of $\mat A$ are these right singular vectors in $\mat V = \mat V_R$. It takes $O(n^3)$ to compute all these PCs from $\mat R$; in contrast, it takes $O(mn^2)$ to compute them directly from $\mat A$.

A truncated SVD of $\mat A$ is $\mat U_{:, 1:k} \mat\Sigma_{1:k, 1:k} \mat V_{:,1:k}^\transpose$,
where we only keep the top-$k$ largest singular values and their corresponding left and right singular vectors. This  defines a $k$-dimensional linear projection of $\mat A$:
\[\mat A \mat V_{:,1:k} = \mat U_{:, 1:k} \mat\Sigma_{1:k, 1:k} \mat V_{:,1:k}^\transpose \mat V_{:,1:k} = \mat U_{:, 1:k} \mat\Sigma_{1:k, 1:k}.\]

Among all possible sets of $k$ $n$-dimensional vectors, the vectors in $\mat V_{:,1:k}$, which are the top-$k$ PCs of $\mat A$, preserve the maximal variance in $\mat A$. Equivalently, they induce the lowest error $||\mat A - \mat A\mat V_{:,1:k}\mat V_{:,1:k}^\transpose||_2$ for reconstructing $\mat A$ as a $k$-dimensional linear projection (Eckart-Young-Mirsky theorem~\cite{Eckart-Young-Mirsky1:PCA:1936,Eckart-Young-Mirsky2:PCA:1960}).
Applications of PCA require the top-$k$ PCs for some value of $k$ and sometimes also the $k$-dimensional linear projection of $\mat A$.

\section{Experiments}
\label{sec:experiments}

\setlength{\tabcolsep}{2pt}
We evaluate the runtime performance and accuracy of \Figaro against Intel MKL 2021.2.0 using the MKL C++ API (MKL called from numpy is slower) and our custom implementations. We also benchmarked OpenBLAS 0.13.3 called from numpy 1.22.0 built from source, but it was consistently slower (1.5x) than MKL, so we do not report its performance further. Both MKL and OpenBlAS implement the Householder algorithm for the QR decomposition of dense matrices~\cite{Householder58b}. %
We also benchmarked THIN as a standalone algorithm to compute $\mat R$ (Sec.\@~\ref{sec:postprocessing}).

We use the following naming conventions: \Figaro-THIN and \Figaro-MKL are \Figaro with THIN and MKL post-proce\-ssing, respectively. In the plots, we also precede the system names by SVD or PCA to denote their versions that compute SVD or PCA respectively. 
\Figaro and THIN use row-major order for storing matrices, while MKL uses column-major order; we verified that they perform best for these orders.

\begin{table}[t]
    \setlength{\tabcolsep}{1.5pt}
        \begin{center}
            \begin{tabular}{|l|r|r|r|r|r|r|}
                \hline
                & \multicolumn{2}{c|}{Retailer (R)} & \multicolumn{2}{c|}{Favorita (F)} & \multicolumn{2}{c|}{Yelp (Y)}\\
                & orig. & OHE & orig. & OHE & orig. & OHE \\
                \hline
                {\bf Input Database} & & & & & & \\
                \# rows (M) & 84 & 0.84 & 125 & 1.28  & 2 & 0.013\\
                size on disk (GB) & 7.9 & 38.9  & 11.7 & 48.4   & 0.52 & 1  \\
                \hline
                {\bf Join Output} & & & & & & \\
                  \# rows (M) & 84 & 0.84 & 127 & 1.27  & 150 & 1.5 \\
                 \# data columns & 43 & 2004 & 30 & 1645  & 50 & 5625\\
                size on disk (GB) & 84.2 & 39.2 & 89.4 & 50 & 175.7 &  197.5 \\
                PSQL time (s) & 143.3 & 0.7 & 217 & 0.8 & 180.8 & 1.5\\
                \hline
            \end{tabular}
        \end{center}
        \caption{Characteristics of the datasets (original) and their one-hot encodings (OHE).}
        \label{table:real-datasets}
    \end{table}

\textbf{Experimental Setup.} All experiments were performed on an Intel Xeon Silver 4214 (24 physical/48 logical cores, 1GHz, 188GiB) with Debian GNU/Linux 10.
We use g++ 10.1 for compiling the C++ code using the Ofast optimization flag.
     The performance numbers are averages over 20 consecutive runs.
We do not consider the time to load the database into RAM and assume that all relations and the join output are sorted by their join attributes. All systems use all cores in all experiments apart from Experiment 2.

\textbf{Datasets.} We use three datasets. Retailer (R) \cite{Schleich:F:2016} and Favorita (F)~\cite{favorita} are used for forecasting user demands and sales. Yelp (Y)~\cite{yelpdataset} has review ratings given by users to businesses. The characteristics of these data\-sets (Table~\ref{table:real-datasets}) are common in retail and advertising, where data is generated by sales transactions or click streams.
Retailer has a snowflake schema, Favorita has a star schema. Both have key-fkey joins, a large fact table, and several small  dimension tables. Yelp has a star schema with many-to-many joins. We also consider a one-hot-encoding version of 1\% of these datasets (OHE), where some keys are one-hot encoded in new data columns. They yield wider matrices. Some plots show performance for a percentage of the join output. The corresponding input dataset is computed by projecting the fragment of the join output onto the schemas of the relations. When taking a percentage of an OHE dataset, we map an attribute domain to that percentage of it using division hashing.

We also use synthetic datasets of relations $\mat{S},\mat T \in \matspace{m}{n}$, whose join is the Cartesian product. The data in each column follows a uniform distribution in the range $[-3, 3)$. For accuracy experiments, we fix (part of) the output $\mat{R}_{\mathit{fixed}}$ and derive the input relation $\mat S$ so that the QR decomposition of the Cartesian product agrees with $\mat{R}_{\mathit{fixed}}$. The advantage of this approach is that $\mat{R}_{\mathit{fixed}}$ can be effectively used as ground truth for checking the accuracy of the algorithms.
\subsection{Summary}

The main takeaway of our experiments is that \Figaro significantly outperforms its competitors in runtime for computing QR, SVD, and PCA, as well as in accuracy for computing $\mat R$, by a factor proportional to the gap between the join output and input sizes.  This factor is up to two orders of magnitude. This includes scenarios with key-fkey joins over real data with millions of rows and tens of data columns and with many-to-many joins over synthetic data with thousands of rows and columns. Whenever the join input and output sizes are very close, however, \Figaro's benefit is small; we verify this by reducing the number of rows and increasing the number of columns of our real data. When the number of data columns is at least the square root of the number of rows, e.g., 1K columns and 1M rows in our experiments, the performance gap almost vanishes. This is to be expected, since the complexity of computing the matrix $\mat R$ increases linearly with the number of rows but quadratically with the number of columns.

We further show that \Figaro's performance depends on the join tree. Also, the accuracy of the orthogonal matrices $\mat Q$ and $\mat U$ computed by \Figaro depends on the condition number of the datasets and can be better or worse than MKL.

Yet there is not one flavor of \Figaro that performs best in all our experiments. In case of thin matrices, i.e., with less than 50 data columns in the join output, or for very sparse matrices, such as those obtained by one-hot encoding, \Figaro-THIN is the best as its QR post-processing phase can exploit effectively the sparsity in the result of \Figaro's Givens rotations. For dense and wide matrices, i.e., with hundreds or thousands of data columns, \Figaro-MKL is the best as its QR post-processing phase uses MKL that parallelizes better than THIN. Therefore, the following experiments primarily focus on the performance (in both runtime and accuracy) of the best of the two algorithms, with brief notes on the performance of the other algorithm. This means that we use \Figaro-MKL for the experiments with the synthetic datasets and \Figaro-THIN for the real-world datasets and their OHE versions.

\Figaro is the only algorithm that works directly on the input database, all others work on the materialized join output. For \Figaro, we report the time to compute both the matrix decompositions and the join {\em intertwined}, whereas for the others we only report the time to compute the matrix decompositions over the {\em precomputed} join matrix. Table~\ref{table:real-datasets} gives the times for materializing the join for the three datasets; these join times are typically larger than the MKL compute times.
\subsection{QR Decomposition}

We first consider the task of computing the upper-triangular matrix $\mat R$ only. After that, we investigate the computation of the orthogonal matrix $\mat Q$.

\begin{figure}[t]
	\includegraphics[width=0.49\textwidth]{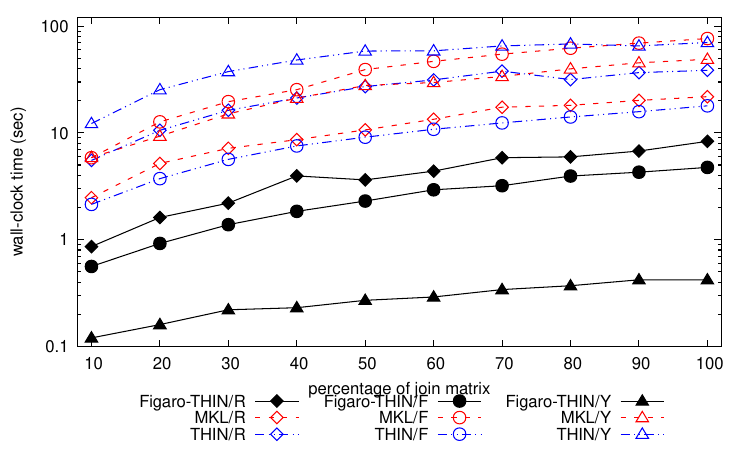}
    \includegraphics[width=0.49\textwidth]{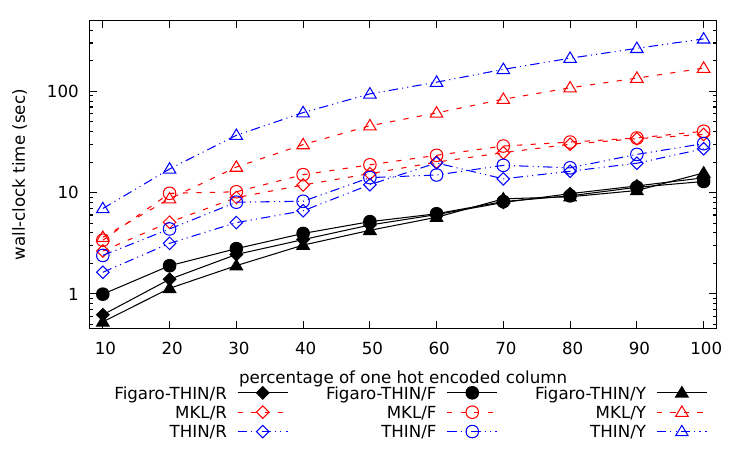}
    \caption{Exp. 1: Runtime performance for computing $\mat R$ in the QR decomposition over the original  datasets R, F, Y (top) and their OHE versions (bottom).}
    \label{fig:experiment1-realworld}
\end{figure}

\textbf{Experiment 1: Runtime performance for computing $\mat R$.} When compared to its competitors, \Figaro performs very well for real data in case the join output is larger than its input. Fig.\@~\ref{fig:experiment1-realworld} (top) gives the runtime performance of \Figaro-THIN, THIN, and MKL for the three real datasets as a function of the percentage of the dataset. The performance gap remains mostly unchanged as the data size is increased.
Relative to MKL, \Figaro-THIN is 2.9x faster for Retailer, 16.1x for Favorita and 120.5x for Yelp.  MKL outperforms THIN, except for Favorita where the latter is 4.3x faster than the former. This is because Favorita has the smallest number of columns amongst the three datasets and THIN works particularly well on thin matrices. We do not show \Figaro-MKL to avoid clutter; it consistently performs worse (up to 3x) than \Figaro-THIN.

We next consider the OHE versions of Retailer and Favorita, for which the join input and output sizes are close. Fig.\@~\ref{fig:experiment1-realworld} (bottom) gives the runtime performance of \Figaro-THIN, THIN, and MKL for the one-hot encoded fragments of the three real datasets as a function of the percentage of one-hot encoded columns.  The strategy of \Figaro to push past joins is less beneficial in this case, as it copies heads and tails along the join tree without a significant optimization benefit.
    Relative to MKL, it is 2.7x faster for Retailer and 3.1x faster for Favorita.
The explanation for this speed-up is elsewhere: Whereas the speed-up in Fig.\@~\ref{fig:experiment1-realworld} (top) is due to structural sparsity, as enforced by the joins, here we have value sparsity due to the many zeros introduced by one-hot encoding. By sorting on the one-hot encoded attribute and allocating blocks with the same attribute value to each thread, we ensure that large blocks of zeros in the one-hot encoding will not be dirtied by the rotations performed by the thread. This effectively preserves many zeros from the input and reduces the number of rotations needed in post-processing to zero values in the output of \Figaro. This strategy needs however to be supported by a good join tree: A relation with one-hot encoded attributes is sorted on these attributes and this order is a prefix of the sorting order used by its join with its parent relation in the join tree.

\begin{figure}[t]
        \begin{center}
            \begin{tabular}{|l||r@{\hspace*{.75em}}r@{\hspace*{.75em}}r@{\hspace*{.75em}}r|}
                \hline
                & \multicolumn{4}{c|}{\#columns}\\
                \hline
                \#rows & $2^6$ & $2^8$ & $2^{10}$ & $2^{12}$\\
                \hline\hline
                $2^9$   & 0.10 &   0.13 &      0.17 &      0.43 \\
                $2^{10}$ & 0.10 &   0.13 &      0.30 &      0.81 \\
                $2^{11}$ & 0.12 &   0.15 &      0.31 &      1.77 \\
                $2^{12}$ & 0.12 &   0.14 &      0.38 &      4.65 \\
                $2^{13}$ & 0.13 &   0.18 &      0.54 &      5.85 \\
                \hline
            \end{tabular}\hspace*{1em}%
            \begin{tabular}{|r@{\hspace*{.75em}}r@{\hspace*{.75em}}r@{\hspace*{.75em}}r|}
                \hline
                \multicolumn{4}{|c|}{\#columns}\\
                \hline
                $2^6$ & $2^8$ & $2^{10}$ & $2^{12}$\\
                \hline\hline
                         3 &     18 &       104 &       368 \\
                        16 &     76 &       246 &       609 \\
                        53 &    278 &       785 &           \\
                        201 &   1056 &           &           \\
                        672 &        &           &           \\
                \hline
            \end{tabular}
        \end{center}

    \caption{Exp. 1: Runtime performance of \Figaro-MKL and MKL for computing $\mat R$ in the QR decomposition of the Cartesian product of two relations. The numbers of rows and columns are per relation; for relations of $2^{13}$ rows columns and $2^{12}$, MKL's input is a $2^{26}\times 2^{13}$ matrix. Left: Runtime performance of \Figaro-MKL (sec). Right: Speed-up of \Figaro-MKL over MKL (rounded to closest natural number). An empty cell means that MKL runs out of memory.}
    \label{fig:experiment1-synthetic}
\end{figure}

We further verified that more input rows allowed \Figaro to scale better, but MKL ran out of memory. THIN and MKL have similar performance. For the Yelp OHE dataset, its size remains much less than its join result and \Figaro outperforms MKL by 11x.

Pronounced benefits are obtained for many-to-many joins, for which the join output is much larger than the input. We verified this claim for the Cartesian products of two relations of thousands of rows and columns. \Figaro outperforms MKL by up to three orders of magnitude on this synthetic data (Fig.\@~\ref{fig:experiment1-synthetic}). As expected, \Figaro scales linearly with the number of rows, while MKL scales quadratically as it works on the materialized Cartesian product.
The speed-up of \Figaro over MKL increases as we increase the number of rows and columns of the two  relations. For wide relations ($2^{12}$ columns), \Figaro-MKL (shown in figure) is up to 10x faster than \Figaro-THIN (not shown).

Most of the time for \Figaro is taken by post-processing. Computing the batch of the group-by aggregates and the tails and heads of the input relations take under 10\% of the overall time for OHE datasets, and about 50\% for the original datasets.

\begin{figure}[t]
    \centering
    \includegraphics[width=0.55\textwidth]{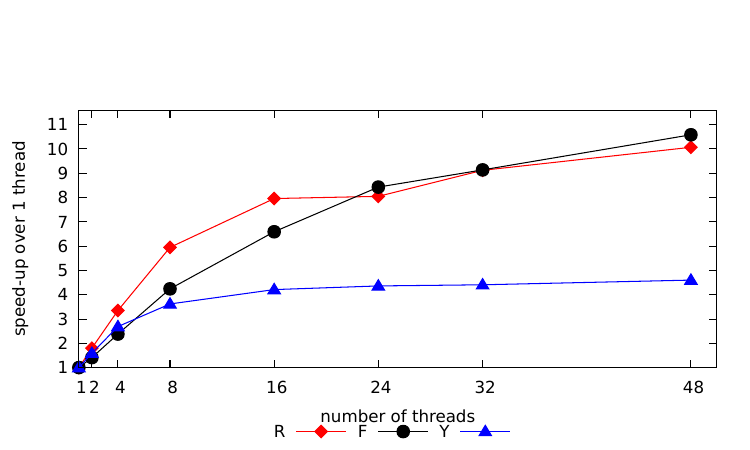}

    \caption{Exp. 2: Speed-up of multi-threading over single-threading for \Figaro-THIN computation of $\mat R$ in the QR decomposition over the three datasets.}
    \label{fig:experiment2-multithreaded}
\end{figure}

\textbf{Experiment 2: Multi-cores scalability.}
 \Figaro uses domain parallelism: It splits each input relation into as many contiguous blocks as available threads and applies the same transformation to each block independently.
 Fig.\@~\ref{fig:experiment2-multithreaded} shows the performance of \Figaro-THIN as we vary the number of threads up to the number of available logical cores (48). The fastest speed-up increase is achieved up to 8 threads for all datasets and up to 16 threads for Retailer and Favorita. The speed-up still increases up to 48 threads, yet its slope gets smaller. This is in particular the case for the smallest dataset Yelp, for which \Figaro-THIN only takes under 0.5 seconds using 8 threads.

\begin{table}[t]
    \begin{center}
        \begin{tabular}{|cc|c|c|c|}
            \hline
            & & Retailer & Favorita & Yelp \\
            \hline
            \multirow{4}{*}{\rotatebox[origin=c]{90}{Original}}
            &\multirow{2}{*}{Bad}&  I(L(C),W,T)  & S(T,(R,O),H,I) & R(B(C,I,H),U) \\
             & & 67.10s  & 29.34s &  1.31s   \\\cline{3-5}
             &\multirow{2}{*}{Good}&  L(C,I(W,T))  & R(T(O,S(H,I))) & B(C,I,H,R(U)) \\
            & & 8.06x  & 6.17x &  3.13x \\
            \hline\hline
            \multirow{4}{*}{\rotatebox[origin=c]{90}{OHE}}
            & \multirow{2}{*}{Bad} &  L(C,{\color{red}I}(W,{T})) &  H({\color{red}S}(T(R,O),{I})) & B(C,I,H,{\color{red}R}(U))  \\
            & & 431.57s & 802.62s & 11.91s \\\cline{3-5}
            & \multirow{2}{*}{Good} & {T}({\color{red}I}(L(C),W))   & {I}({\color{red}S}(T(R,O),H))&   H(B({\color{red}R}(U),C,I)) \\
            &  & 21.25x  & 47.13x & 1.32x \\
            \hline
        \end{tabular}
    \end{center}
    \caption{Exp.\@~3: \Figaro-THIN's runtime to compute $\mat R$ in the QR decomposition using a bad join tree and relative speed-up using a good join tree. The join trees use term notation over the abbreviated relation names:  Retailer: Inventory(I), Item(T) , Weather(W), Location(L), Census(C); Favorita: Sales(S), Stores(R), Oil(O), Holidays(H), Items(I), Transactions(T); Yelp: Business(B), Category(C), CheckIn(I), User(U), Hours(H), Review(R). The relation with OHE values is in red.}
    \label{tab:join-orders}
\end{table}

\textbf{Experiment 3: Effect of join trees.} The runtime of \Figaro is influenced by the join tree.
    As in classical query optimization, \Figaro prefers join trees such that the overall size of the intermediate results is the smallest.
    For our datasets, this translates to having the large fact tables involved earlier in the joins.
We explain for Retailer, whose large and narrow fact table Inventory uses key-fkey joins on composite keys (item id, date, location) to join with the small dimension tables Locations, Weather, Item and Census. \Figaro aggregates away join keys as it moves from leaves to the root of the join tree. By aggregating away join keys over the large table as soon as possible, it creates small intermediate results and reduces the number of copies of dimension-table tuples in the intermediate results to pair with tuples in the fact table. If the fact table would be the root, then \Figaro would first join it with its children in the join tree and then apply transformations, with no benefit over MKL as in both cases we would work on the materialized join output. Therefore, a bad join tree for Retailer has Inventory (I) as a root. In contrast, a good join tree first joins Inventory with Weather and Item, thereby aggregating away the item id and date keys and reducing the size of the intermediate result to be joined with Census and Location.
    As shown in Table~\ref{tab:join-orders}, \Figaro-THIN performs 8.06x faster using a good join tree instead of a bad one.

For the OHE datasets, a key performance differentiator is the sorting order of the relations with one-hot encoded attributes, as explained in Experiment 1. For instance, the relation Inventory in the Retailer dataset (colored red in Table~\ref{tab:join-orders}) has the one-hot encoded attribute product id. In a bad join tree, it is not sorted on this attribute: it joins with its parent relation Location on location id, so it is primarily sorted on location id. In a good join tree, Inventory is sorted on product id as it joins with its parent Item on product id.

\begin{figure}[t]
	\includegraphics[width=0.49\textwidth]{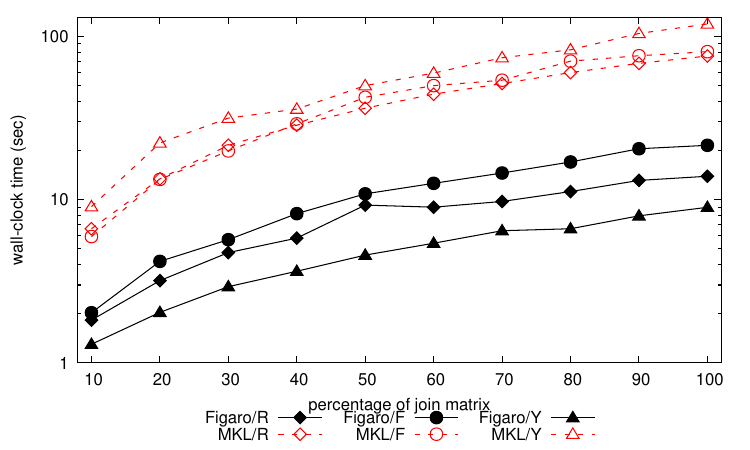}
    \includegraphics[width=0.49\textwidth]{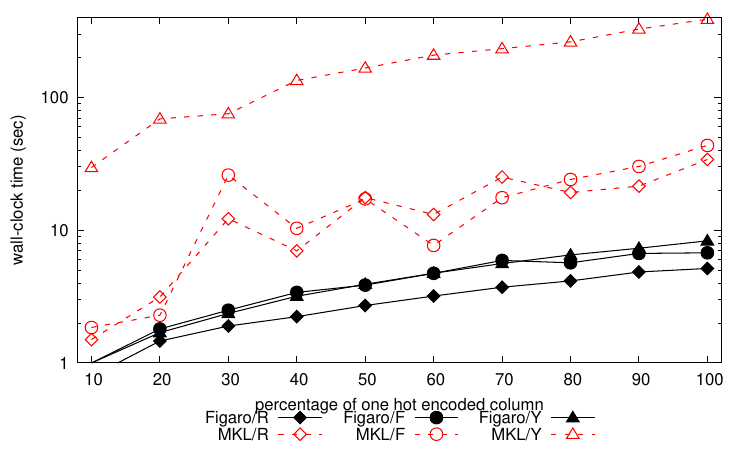}
    \caption{Exp. 4: Runtime performance for computing the fully materialized $\mat Q$ for the original  datasets R, F, Y (top) and their OHE versions (bottom).}
    \label{fig:experiment1-realworld_q}
\end{figure}

\textbf{Experiment 4: Runtime performance for computing $\mat Q$.}
\Figaro outperforms MKL for computing the orthogonal matrix $\mat Q$ (Fig.\@~\ref{fig:experiment1-realworld_q}). For this experiment, \Figaro takes as input the already computed $\mat R$, while MKL takes as input the precomputed Household vectors, which are also based on $\mat R$.
The speed-up is 5.4x for Retailer, 3.7x for Favorita, and 13x for Yelp (Fig.\@~\ref{fig:experiment1-realworld_q} top). As we linearly increase the join matrix size, the runtime performance of \Figaro and MKL increases linearly.
For the OHE datasets, the speed-up is 1.9-6.7x for Retailer, 1.9-10.5x for Favorita, and 29-47x for Yelp (Fig.\@~\ref{fig:experiment1-realworld_q} bottom).

\begin{figure}[t]
	\includegraphics[width=0.49\textwidth]{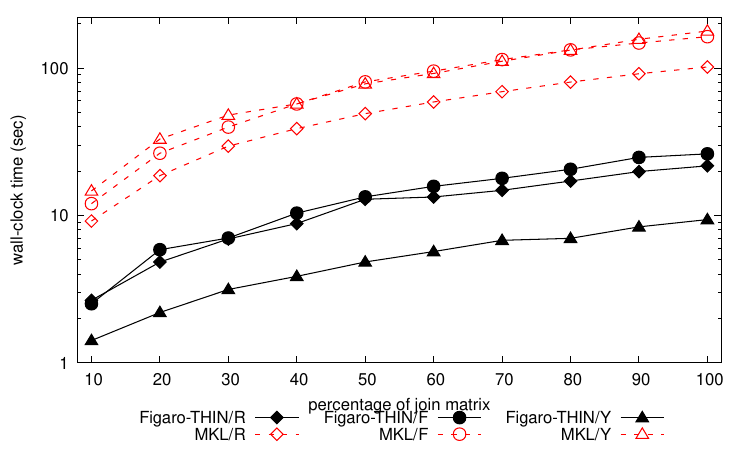}
    \includegraphics[width=0.49\textwidth]{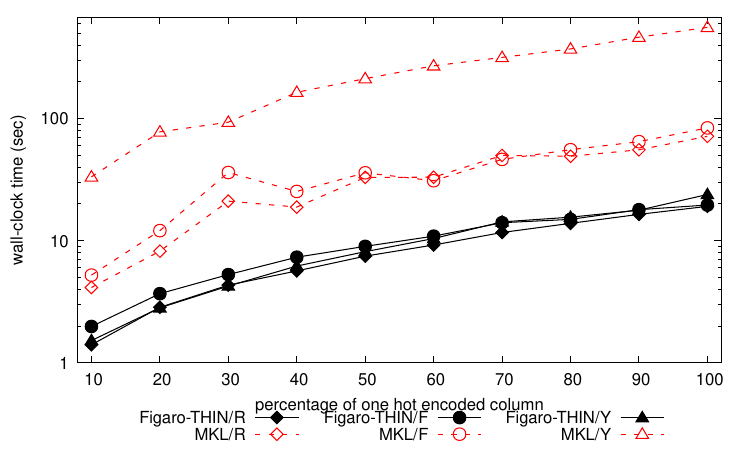}
    \caption{Exp. 5: Runtime performance for end-to-end QR decomposition for the original  datasets R, F, Y (top) and their OHE versions (bottom).}
    \label{fig:experiment1-realworld_qr}
\end{figure}

\textbf{Experiment 5: Runtime performance for end-to-end QR decomposition.} To complete the picture, Fig.\@~\ref{fig:experiment1-realworld_qr} reports the performance for computing both  $\mat R$ and $\mat Q$. The input to \Figaro-THIN is the database and to MKL is the materialized join output. As expected, \Figaro-THIN clearly outperforms MKL for all considered datasets.

\textbf{Experiment 6: Accuracy of computed $\mat R$.} To assess the accuracy of computing $\mat R$, we designed a synthetic dataset consisting of two relations and want to compute the upper-triangular $\mat R$ in the QR decomposition of their Cartesian product. The input relations are defined based on a given matrix $\mat{R}_{\mathit{fixed}}$ such that $\mat{R}_{\mathit{fixed}}$ is part of  $\mat R$. The construction is detailed in App.\@~\ref{sec:accuracy-design}.
We report the relative error $\frac{\pnorm{\mat{R}_{\mathit{fixed}} - \hat{\mat{R}}_{\mathit{fixed}}}{F}}{\pnorm{\mat R_{\mathit{fixed}}}{F}}$ of the computed partial result $\hat{\mat{R}}_{\mathit{fixed}}$ compared to the ground truth $\mat{R}_{\mathit{fixed}}$, where $\pnorm{\cdot}{F}$ is the Frobenius norm.

\begin{table}[t]
    \begin{center}
        \begin{tabular}{|l||r@{\hspace*{.75em}}r@{\hspace*{.75em}}r|}
            \hline
            & \multicolumn{3}{c|}{\#columns}\\
            \hline
            \#rows  & $2^{4}$ & $2^{6}$ & $2^8$\\
            \hline
            \hline
            $2^9$   & 2.3e-15 & 1.8e-14 & 3.7e-14 \\
            $2^{10}$ & 3.5e-15 & 3.3e-14 & 1.3e-13 \\
            $2^{11}$ & 4.7e-15 & 4.3e-14 & 3.2e-13 \\
            $2^{12}$ &   6e-15 & 5.4e-14 & 5.2e-13 \\
            $2^{13}$ & 7.9e-15 & 6.3e-14 &         \\
            \hline
            \end{tabular}\hspace*{.5em}%
            \begin{tabular}{|r@{\hspace*{.75em}}r@{\hspace*{.75em}}r|}
            \hline
            \multicolumn{3}{|c|}{\#columns}\\
            \hline
              $2^{4}$ & $2^{6}$ & $2^8$ \\
            \hline
            \hline

            26 &     1.5 &    1.2  \\
            73 &     5.3 &    1.3 \\
            250 &      20 &    1.6 \\
            830 &      64 &    5.4 \\
            2600 & 250 &        \\
            \hline
            \end{tabular}

    \end{center}
    \caption{Exp. 6: (Left) Error for \Figaro-MKL relative to the ground truth. (Right) Division of relative error of MKL over the relative error of \Figaro-MKL. The empty bottom-right cell is due to the out-of-memory error for MKL.}
    \label{tab:figaro_thin_relative_part_error}
\end{table}

Table~\ref{tab:figaro_thin_relative_part_error} (left) shows the error of \Figaro-MKL as we vary the number of rows and columns of the two relations following a geometric progression.
As the number of rows increases, the accuracy only changes slightly. The accuracy drops as the number of columns increases. The error remains however sufficiently close to the machine representation unit ($10^{-16}$). We also verified that \Figaro-THIN's error is very similar.

\begin{figure}[t]
	\includegraphics[width=0.49\textwidth]{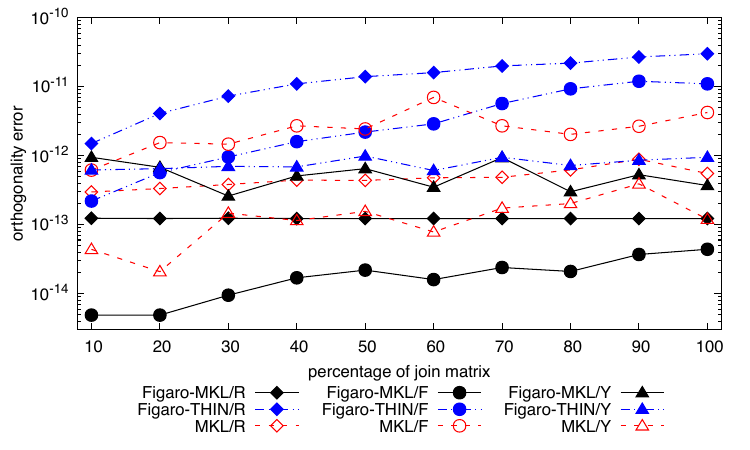}
    \includegraphics[width=0.49\textwidth]{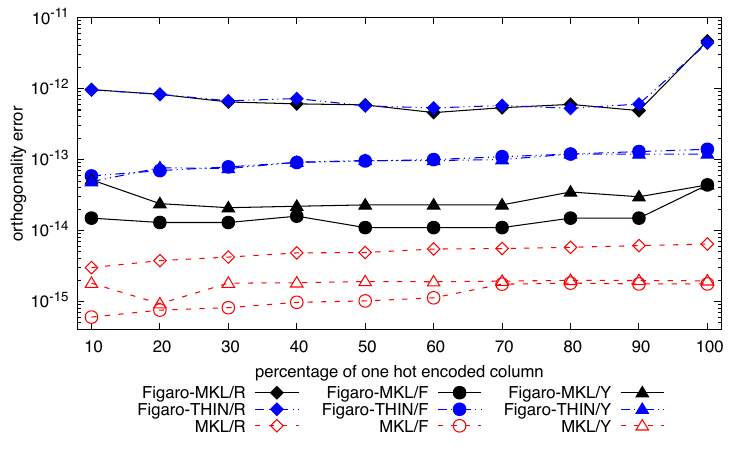}
    \caption{Exp. 7: Orthogonality error of $\mat{Q}$ computed by both \Figaro variants and MKL for the three datasets (top) and their OHE versions (bottom). }
    \label{fig:experiment6-qr-ort}
\end{figure}

We also computed the relative error for MKL. Table~\ref{tab:figaro_thin_relative_part_error} (right) shows the result of dividing the relative errors of MKL and \Figaro-MKL. A number greater than $1$ means that \Figaro-MKL is more accurate than MKL. The error gap increases with the number of rows and decreases with the number of columns. The latter is as expected, as post-processing dominates the computation for wide matrices. \Figaro-MKL is up to three orders of magnitude more accurate than MKL.

\textbf{Experiment 7: Accuracy of computed $\mat Q$.}
The accuracy of computing $\mat Q\in \matspace{m}{n}$ is given by the orthogonality error $\frac{\frobnorm{{\matD{Q}^{\transpose}} \mat{Q} - \matD{I}}}{\frobnorm{\mat{I}}}$, where $\mat{I} \in \matspace{n}{n}$ is the identity matrix and $\frobnorm{\cdot}$ is the Frobenius norm~\cite[p.~360]{Higham}. An orthogonality error of 0 is the ideal outcome, it means that $\mat Q$ is perfectly orthogonal.

Fig.\@~\ref{fig:experiment6-qr-ort} depicts the orthogonality errors of the matrices $\mat Q$ constructed by \Figaro-MKL, \Figaro-THIN and MKL. We disregard \Figaro-THIN in the following analysis, as it is less  accurate than \Figaro-MKL in this experiment. This is as expected, as the THIN post-processing is far less optimized with respect to numerical stability than MKL. We see that \Figaro-MKL is more accurate for Retailer (one order of magnitude) and Favorita (two orders of magnitude), but less accurate for Yelp (one order of magnitude).
The reason is that the \emph{condition number}~\cite{Higham} of the join matrix $\mat A$ influences the orthogonality of $\mat{Q}$ computed by \Figaro. The condition number is the largest singular value of $\mat A$ divided by the smallest singular value of $\mat A$. A very large condition number means that the smallest singular value is very close to $0$, so, the matrix is almost singular~\cite{rump1997bounds,demmel1987condition}. If this is the case for $\mat A$, the same holds for the matrix $\mat R$ of its QR decomposition, as they have the same singular values. Recall that \Figaro inverts $\mat R$ to compute $\mat Q$ (Sec.\@~\ref{sec:beyondR}), this step may introduce larger rounding errors when $\mat R$ is close to being singular. %

\begin{table}[t]
    \begin{center}
        \begin{tabular}{|l||l|l|l|}
        \hline
        Measurement                                          & Dataset  & Real     & OHE       \\ \hline
        \multirow{3}{*}{Condition number} & Retailer & 3.75E+07 & 2.72E+09   \\
                                        & Favorita & 6.90E+04 & 3.41E+06   \\
                                        & Yelp     & 1.70E+10 & 1.78E+10   \\ \hline \hline
        \multirow{3}{*}{Smallest singular value}                & Retailer & 36.2052  & 0.047772  \\
                                                            & Favorita & 482.358  & 0.367022  \\
                                                            & Yelp     & 2.19208  & 0.208876  \\ \hline
        \end{tabular}
    \end{center}
        \caption{Exp. 7: Condition numbers and smallest singular values for the three datasets and their OHE versions.}
    \label{tab:cond_nums}
\end{table}

\begin{table}[t]
    \begin{center}
      \begin{tabular}{|l||r@{\hspace*{.75em}}r@{\hspace*{.75em}}r|}
            \hline
            & \multicolumn{3}{c|}{\#columns}\\
            \hline
            \#rows  & $2^{4}$ & $2^{6}$ & $2^8$\\
            \hline
            \hline
            $2^9$   & 1.3E-14 & 1.7E-14 & 3.9E-14 \\
            $2^{10}$ &   6.2E-15 & 2.1E-14 & 5.7E-13 \\
            $2^{11}$ & 7.0E-14 &  7.3E-14 & 1.3E-13 \\
            $2^{12}$ & 2.8E-14 & 4.1E-14 & 1.5E-13 \\
            $2^{13}$ & 2.2E-13 & 2.0E-13 &         \\
            \hline
            \end{tabular}\hspace*{.5em}%
            \begin{tabular}{|r@{\hspace*{.75em}}r@{\hspace*{.75em}}r|}
            \hline
            \multicolumn{3}{|c|}{\#columns}\\
            \hline
              $2^{4}$ & $2^{6}$ & $2^8$ \\
            \hline
            \hline

            3.7 &     1.9 &    0.9  \\
            21 &     8.8 &    2.3 \\
            9.8 &      11.9 &    3.9 \\
            112.1 &      93.1 &    18.6 \\
            56.1 & 81.0 &        \\
            \hline
        \end{tabular}

    \end{center}
    \caption{Exp. 7: (Left) Orthogonality error of $\mat{Q}$ for \Figaro-MKL over the Cartesian product. (Right) Division of orthogonality error of MKL by the orthogonality error of \Figaro-MKL. The right-bottom cell is empty as MKL ran out of memory.}
    \label{tab:accuracy_syn_ort_mkl_figaro}
\end{table}

Table \ref{tab:cond_nums} gives the condition numbers and smallest singular values for the join matrix $\mat A$ for each of our three datasets and their OHE versions. Since Favorita has the smallest condition number, the orthogonality error of $\mat{Q}$ computed by \Figaro-MKL is very low, i.e., $\mat{Q}$ is very close to an orthogonal matrix. In contrast, Yelp has the largest condition number and \Figaro produces a less orthogonal matrix $\mat{Q}$ for this dataset.

Table \ref{tab:cond_nums} also shows that the condition numbers are much larger for the OHE versions of Retailer and Favorita (by a factor of 100x), while the condition number remains almost the same for the OHE version of Yelp.  We therefore expect less accurate matrices $\mat Q$ computed by \Figaro-MKL. This is indeed the case, as shown in Fig.\@~\ref{fig:experiment6-qr-ort} (bottom). Remarkably, MKL improves the orthogonality of $\mat Q$ in case of the OHE datasets relative to the original datasets, which suggests that it can effectively exploit the sparsity due to the one-hot encoding to avoid accumulating too many rounding errors.

Table~\ref{tab:accuracy_syn_ort_mkl_figaro} gives the orthogonality error of $\mat Q$ for the Cartesian product of two relations as we vary their numbers of rows and columns. The condition numbers, the smallest singular values, and even the orthogonality vary in the same range as for Retailer original and OHE in Table~\ref{tab:cond_nums}. The orthogonality error for $\mat Q$ computed by \Figaro-MKL remains rather low and is 1-100x lower than for MKL (Table~\ref{tab:accuracy_syn_ort_mkl_figaro} right). The gap between the two systems closes as we increase the number of columns.

\subsection{Singular Value Decomposition}

We extended \Figaro to compute the SVD of the join matrix $\mat A$ as detailed in Sec.\@~\ref{sec:beyondR}. We call this extension \svdFigaro. It works directly on the input database.
We compare it against \svdMKL, the divide\&conquer approach of MKL that computes the QR decomposition of $\mat A$ and then bidiagonalizes the upper-triangular matrix $\mat R$.
When only singular values are computed, all methods use the dqds algorithm~\cite{fernando1994accurate} to compute these values from the intermediate bidiagonal matrices.
We also experimented with the QR iteration, the power iteration, and the eigendecomposition approaches. QR iteration performs similar to \svdMKL, although for the OHE datasets it is one order of magnitude less accurate, where accuracy is measured as the orthogonality of the matrix $\mat U$. Power iteration is at least one order of magnitude slower than \svdFigaro for achieving a comparable accuracy. In our tests, the orthogonality error of the eigendecomposition approach was much larger than for the other approaches.

\begin{figure}[t]
	\includegraphics[width=0.49\textwidth]{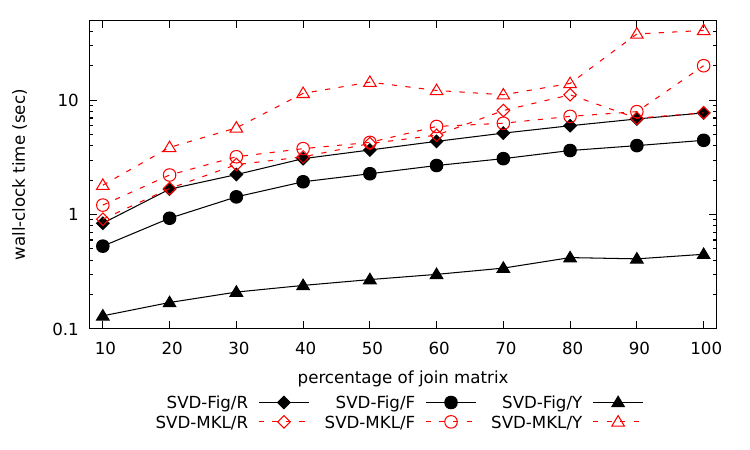}
    \includegraphics[width=0.49\textwidth]{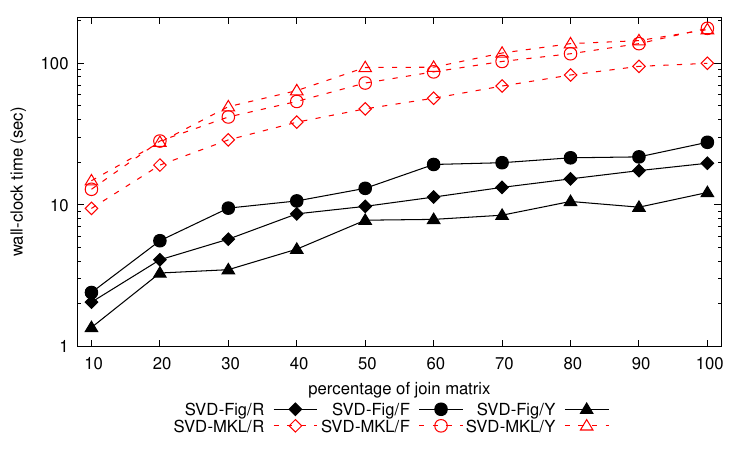}
    \caption{Exp. 8: Runtime performance of \svdFigaro and MKL over the original datasets R, F and Y. Top: Only singular values ($\mat \Sigma$) are computed. Bottom: The entire SVD is computed: $\mat{U}, \mat{\Sigma}, \mat{V}$.}
    \label{fig:experiment7-realworld-svd}
\end{figure}

\begin{figure}[t]
    \includegraphics[width=0.49\textwidth]{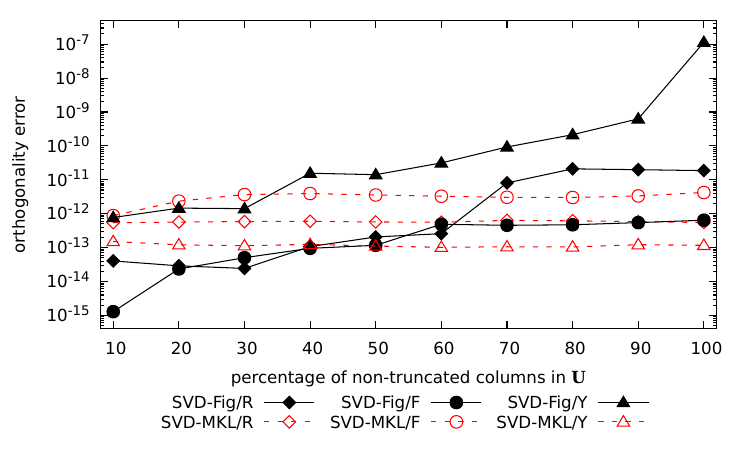}
    \includegraphics[width=0.49\textwidth]{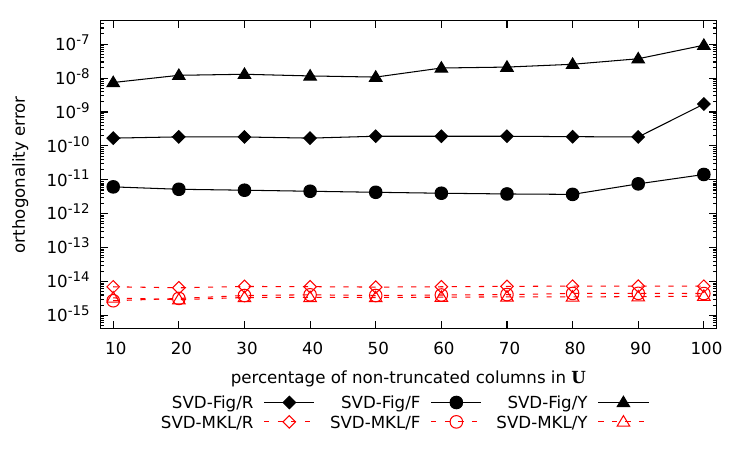}
    \caption{Exp. 9: Orthogonality error for $\mat{U}$ computed by \svdFigaro and MKL for our datasets R, F and Y as we vary the percentage of non-truncated columns in $\mat{U}$. Top: Original datasets. Bottom: OHE versions.}
    \label{fig:experiment8-realworld-svd-ort}
\end{figure}

\textbf{Experiment 8: Runtime performance for computing SVD.}
Fig.\@~\ref{fig:experiment7-realworld-svd} (bottom) shows that \svdFigaro clearly outperforms \svdMKL. They both scale linearly with the join output size, yet \svdFigaro is 5x faster than \svdMKL for Retailer and Favorita and 16x for Yelp.
When only the singular values are computed (Fig.\@~\ref{fig:experiment7-realworld-svd} top), \svdFigaro outperforms \svdMKL  by factors 1-1.9x for Retailer, 2-4.5x for Favorita, and up to 90x for Yelp. The reason for the large gap for Yelp is twofold. The singular values of $\mat A$ are the same as for the upper-triangular matrix $\mat R$ in the QR decomposition of $\mat A$. To compute $\mat R$, \Figaro does not need the materialized join matrix $\mat A$, which is much larger than the Yelp input dataset.
For the OHE  datasets, \svdFigaro is faster than \svdMKL by 2-3x faster for Retailer, 2-5x for Favorita, and 16-22x for Yelp (not shown in the figure).
These speed-ups are at the same scale as in Experiment 4.

\textbf{Experiment 9: Accuracy of computed SVD.}
We first investigate the orthogonality error for the truncated matrix $\mat{U}$ of left singular vectors as we vary the percentage of non-truncated columns (Fig.\@~\ref{fig:experiment8-realworld-svd-ort}).
The reason we look into the accuracy of truncated $\mat U$ is twofold. First, this is the largest matrix in the SVD of $\mat A$. Second, its truncated version is used for PCA.
When the percentage of non-truncated columns increases, the orthogonality error increases for \svdFigaro, for MKL it remains roughly the same.
This is because the condition number of the join matrix $\mat A$ grows with the number of non-truncated columns.

\begin{figure}[t]
\centering
    \includegraphics[width=0.55\textwidth]{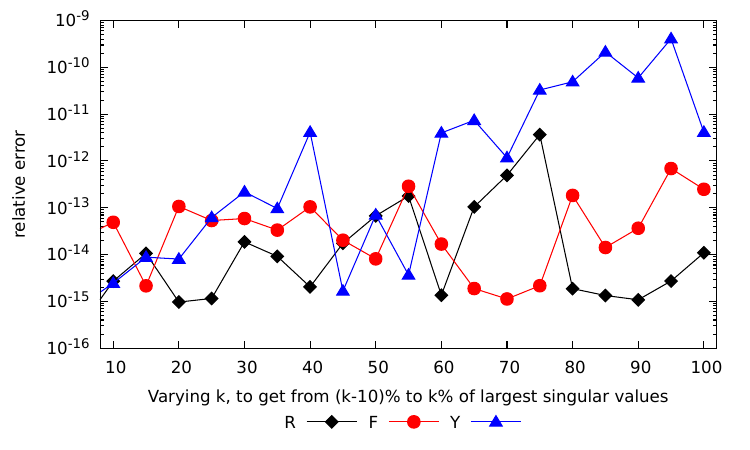}
    \caption{Exp. 9:  Error of \svdFigaro relative to MKL for computing the top $k-5$ to $k$ percent of the largest singular values of the join matrices for the original datasets.}
    \label{fig:experiment8-realworld-svd-sing-vals}
\end{figure}

We see a similar behaviour for the accuracy of the computed singular values.
Fig.\@~\ref{fig:experiment8-realworld-svd-sing-vals} reports the relative error of a sliding window of 5\% of the singular values as computed by $\svdFigaro$ and by MKL, sliding over all singular values in decreasing order. For vectors $\mat v_1$ and $\mat v_2$ computed by $\svdFigaro$ and respectively MKL, the relative error is $\frac{\frobnorm{\mat v_1-\mat v_2}}{\frobnorm{\mat v_2}}$.
The difference between the singular values as computed by \svdFigaro and MKL tends to be small for the largest singular values and increases for smaller singular values.
Also, the singular values computed by \svdFigaro are closer to the values computed by MKL for the join matrices with smaller condition numbers such as Favorita, while for Retailer and Yelp they are more different. We observed similar behaviour for the OHE datasets.

We finally investigate the accuracy for the synthetic dataset and considered different percentages of truncated columns in the matrix $\mat{U}$.
Table~\ref{tab:accuracy_syn_ort_40_svd_mkl_figaro} reports the orthogonality error
of the matrix $\mat{U}$ truncated to 40\% of columns. The matrix $\mat{U}$ computed by \svdFigaro is more orthogonal than the one computed by MKL, the reasoning is similar to the orthogonality of $\mat{Q}$ in Experiment~7.
If we compute the entire matrix $\mat U$, then \svdFigaro incurs a higher orthogonality error than MKL due to the large condition numbers, as shown in Table~\ref{tab:accuracy_syn_ort_100_svd_mkl_figaro}.

\begin{table}[t]
    \begin{center}
      \begin{tabular}{|l||r@{\hspace*{.75em}}r@{\hspace*{.75em}}r|}
            \hline
            & \multicolumn{3}{c|}{\#columns}\\
            \hline
            \#rows  & $2^{4}$ & $2^{6}$ & $2^8$\\
            \hline
            \hline
            $2^9$ & 1.8E-14& 2.7E-15& 6.1E-15\\
            $2^{10}$&  3.1E-15& 2.9E-15& 7.2E-15\\
            $2^{11}$ & 1.8E-15& 1.1E-14& 7.9E-15\\
            $2^{12}$ & 2.4E-15& 7.7E-15& 6.3E-15\\
            $2^{13}$ & 3.9E-15& 5.4E-15 & \\

            \hline
            \end{tabular}\hspace*{.5em}%
            \begin{tabular}{|r@{\hspace*{.75em}}r@{\hspace*{.75em}}r|}
            \hline
            \multicolumn{3}{|c|}{\#columns}\\
            \hline
              $2^{4}$ & $2^{6}$ & $2^8$ \\
            \hline
            \hline
                0.3& 2.5& 1.8 \\
                3.9& 4.8& 4.1 \\
                12.9& 2.7& 6.9  \\
                21.7& 8.0& 29.0 \\
                25.9& 25.1&\\
            \hline
        \end{tabular}

    \end{center}
    \caption{Exp. 9: (Left) Orthogonality error for $\mat U$ computed by \svdFigaro{} and truncated to  40\% columns. (Right) Division of orthogonality error of MKL by the orthogonality error of \svdFigaro{}, the result greater than one means \svdFigaro{} is more accurate. The right-bottom cell is empty as MKL ran out of memory.}
    \label{tab:accuracy_syn_ort_40_svd_mkl_figaro}
\end{table}

\begin{table}[t]
    \begin{center}
      \begin{tabular}{|l|r@{\hspace*{.5em}}r@{\hspace*{.5em}}r|}
            \hline
            & \multicolumn{3}{c|}{\#columns}\\
            \hline
            \#rows  & $2^{4}$ & $2^{6}$ & $2^8$\\
            \hline
            \hline
            $2^9$ & 4.2E-12 & 6.3E-12 &4.2E-11\\
            $2^{10}$ & 9.5E-12 &6.5E-12&4.5E-11\\
            $2^{11}$ & 1.5E-12&1.1E-11&5.0E-11\\
            $2^{12}$ & 1.2E-11&3.0E-11&6.1E-11\\
            $2^{13}$ & 2.0E-11&3.7E-11&\\

            \hline
            \end{tabular}\hspace*{.2em}%
            \begin{tabular}{|r@{\hspace*{.5em}}r@{\hspace*{.5em}}r|}
            \hline
            \multicolumn{3}{|c|}{\#columns}\\
            \hline
              $2^{4}$ & $2^{6}$ & $2^8$ \\
            \hline
            \hline
                1.3E-03&5.0E-03&8.7E-04\\
                1.4E-02&2.9E-02&2.9E-03\\
                4.5E-01&8.0E-02&9.9E-03\\
                2.7E-01&1.3E-01&4.7E-02\\
                6.3E-01&4.3E-01&\\
            \hline
        \end{tabular}

    \end{center}
    \caption{Exp. 9: (Left) Orthogonality error for the full $\mat U$ computed by \svdFigaro{}. (Right) Division of the orthogonality error of MKL by the orthogonality error of \svdFigaro{}, the result greater than one means \svdFigaro{} is more accurate. The right-bottom cell is empty as MKL ran out of memory.}
    \label{tab:accuracy_syn_ort_100_svd_mkl_figaro}
\end{table}

\subsection{Principal Component Analysis}

Sec.\@~\ref{sec:beyondR} shows how to compute PCA using SVD. The top-$k$ principal components of the join matrix $\mat A$ are the right singular vectors in $\mat V$ that correspond to the top-$k$ largest singular values of $\mat A$. Both $\mat V$ and the singular values of $\mat A$ are also of the upper-triangular matrix $\mat R$ computed by \Figaro. The projection of $\mat A$ onto the $k$-dimensional space is given by $\mat U_{:,1:k} \mat \Sigma_{1:k,1:k}$, where $\mat U_{:,1:k}$ is the truncated matrix $\mat U$ of left singular vectors and the diagonal matrix $\mat \Sigma_{1:k,1:k}$ has the $k$ largest singular values along the diagonal. Experiment 9 on the accuracy of the computed truncated $\mat U$ and the singular values carry over to PCA immediately.

\begin{figure}[t]
\centering
	\includegraphics[width=0.52\textwidth]{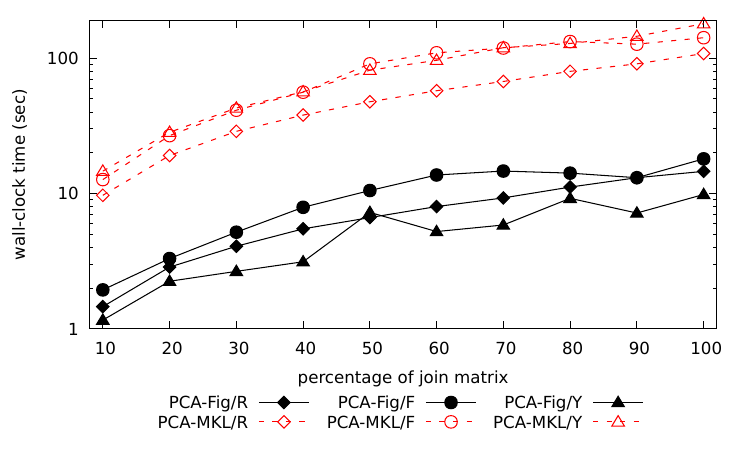}
    \caption{Exp. 10: Runtime performance of \pcaFigaro and MKL for our datasets R, F, and Y.}
    \label{fig:experiment9-realworld-pca}
\end{figure}

\textbf{Experiment 10: Runtime performance for co\-mputing PCA.} Fig.\@~\ref{fig:experiment9-realworld-pca} reports the runtimes to compute $\mat U_{:,1:k} \mat \Sigma_{1:k,1:k}$ for $k= 10\%$ of the number of data columns, i.e, $k$ is 4 for Retailer, 3 for Favorita, and 5 for Yelp. \pcaFigaro is the \Figaro adaptation  to PCA. It takes as input the database. \pcaMKL is a custom implementation that uses MKL to compute SVD from the join matrix $\mat A$ and then multiplies the truncated matrices $\mat{U}$ and $\mat \Sigma$.
For this experiment, we did not center the data. Fig.\@~\ref{fig:experiment9-realworld-pca} shows a consistent gap of one order of magnitude between the two systems. This is similar to the performance reported in Fig.\@~\ref{fig:experiment7-realworld-svd}, since the most expensive computation is taken by the SVD.

\Figaro can compute the principal components corresponding to the largest singular values in our experiments faster than MKL and with similar accuracy. \Figaro is thus a good alternative to MKL in case of matrices defined by joins over relational data.

\section{Related Work}
\label{sec:related}

Our work sits at the interface of linear algebra and databases and is the first to investigate QR decomposition over database joins using Givens rotations. It  complements seminal seven-decades-old work on QR decomposition of matrices. Our motivation for this work is the emergence of machine learning algorithms and applications that are expressed using linear algebra and are computed over relational data.

{\bf Improving performance of linear algebra operations.} Factorized computation~\cite{Olteanu:FDB:2016} is a paradigm that uses structural compression to lower the cost of computing a variety of database and machine learning workloads. It has been used for the efficient computation of $\mat A^{\mat T}\mat A$ over training datasets $\mat A$ created by feature extraction queries, as used for learning a variety of linear and non-linear regression models~\cite{Schleich:F:2016,LiC019,lmfao,SparseTensor:TODS:2020,Olteanu:Learning:2020}. It is also used for linear algebra operations, such as matrix multiplication and element-wise matrix and scalar operations~\cite{ChenKNP17}. Our work also uses a factorized multiplication operator of the non-materialized matrix $\mat A$ and another materialized matrix for the computation of the orthogonal matrix $\mat Q$ in the QR decomposition of $\mat A$ and for the SVD and PCA of $\mat A$.
Besides structural compression, also value-based compression is useful for fast in-memory matrix-vector multiplication~\cite{ElgoharyBHRR19}. Existing techniques for matrix multiplication, as supported by open-source systems like SystemML~\cite{Boehm:SystemML:2014}, Spark~\cite{SparkMatrix:SIGKDD:2016}, and Julia~\cite{Julia}, exploit the sparsity of the  matrices. This calls for sparsity estimation in the form of zeroes~\cite{Sommer0ERH19}.

The matrix layout can have a large impact on the performance of matrix computation in a distributed environment~\cite{LuoJYJ21}. Distributed database systems that can support linear algebra computation~\cite{LuoGGPJJ20} may represent an alternative to high-performan\-ce computing efforts such as ScaLAPACK~\cite{ScaLAPACK:1995}. There are distributed algorithms for QR decomposition in ScaLAPACK~\cite{QR:ScaLAPACK:2014,QR:ScaLAPACK:2019}.

{\bf QR decomposition.} There are three main approa\-ches to QR decomposition of a materialized matrix $\mat A$. The first one is based on {\em Givens rotations}~\cite{Givens58}, as also considered in our work.
Each application of a Givens rotation to a matrix $\mat A$ sets one entry of $\mat A$ to $0$. This method can be parallelized, as each rotation only affects two rows of $\mat A$. It is particularly efficient for sparse matrices. One form of sparsity is the presence of many zeros. We show its efficiency for another form of sparsity: the presence of repeating blocks whose one-off transformation can be reused multiple times.

\emph{Householder transformations}~\cite{Householder58b} are particularly efficient for dense matrices~\cite[p.~366]{Higham}. MKL and openblas used in our experiments implement this method. One Householder transformation sets all but one entry of a vector to $0$, so the QR decomposition of an $m \times n$ matrix can be obtained using $n$ transformations. Both Givens and Householder are numerically stable~\cite[Chapter~19]{Higham}.

The \emph{Gram-Schmidt process}~\cite{Gram83,Schmidt07} computes the orthogonal co\-lumns of $\mat Q$ one at a time by subtracting iteratively from the $i$-th column of $\mat A$ all of its projections onto the  previously computed $i-1$ orthogonal columns of $\mat Q$. A slightly modified variant~\cite{Bjorck67} is numerically stable. The modified Gram-Schmidt is mathematically and numerically equivalent to applying Householder transformations to a matrix that is padded with zeros~\cite{BjorckP92}.

{\bf Implementing Givens rotations.} There are several options on how to set the values $c$ and $s$ of a Givens rotation matrix in the presence of negative values. We follow the choice proposed by Bindel~\cite{BindelDKM02}.
Anderson~\cite{lawn150} defends the original choice~\cite{Givens58} of signs using numerical stability arguments.
Gentleman~\cite{Gentleman73} shows that one can compute an upper-triangular matrix $\mat R'$ and a diagonal matrix $\mat D$ without computing square roots such that $\mat R = \mat D^{\frac{1}{2}} \mat R'$. Stewart~\cite[p.~291]{Stewart} discusses in depth these fast rotations.

{\bf SVD. }
The approach to computing the SVD that is closest to ours proceeds as follows \cite[p.~285]{matrix2016comp}, \cite{lawson1995solving}. It computes the matrices $\mat R$ and $\mat Q$ in the QR decomposition of $\mat{A}$, followed by the computation of the SVD of $\mat{R} = \mat U_R \mat \Sigma_R \mat V_R^\transpose$. Finally, it computes the orthogonal matrix $\mat U = \mat Q \mat U_R$. The SVD of $\mat A$ is then $\mat U \mat \Sigma_R \mat V_R^\transpose$. Our approach  does not compute $\mat Q$ but instead expresses it as a multiplication of the non-materialized $\mat A$ and the inverse of $\mat R$. Further approaches use Householder transformations  to transform $\mat A$ into a bidiagonal matrix \cite[p.~284]{matrix2016comp}, \cite{golub1965calculating}, or compute the eigenvalue decomposition of the matrix $\mat{A}^{\transpose}\mat{A}$ \cite[p.~ 486]{matrix2016comp}, where the singular values are the square roots of the eigenvalues and the matrix $\mat{V}$ consists of the corresponding eigenvectors. All prior approaches require the input matrix $\mat A$ to be materialized.
The existing approaches to computing an SVD of a bidiagonal matrix vary in terms of accuracy~\cite{cline2006computation}. When only singular values are required, the squareroot-free or dqds method is the most popular~\cite{rutishauser1954quotienten,fernando1994accurate}.
The QR-iteration~\cite{demmel1990accurate,demmel1994faster} and divide-and-conquer methods~\cite{gu1995divide} are used in case also the left and right singular matrices are required.

{\bf PCA. } Principal component analysis is a technique for reducing the dimensionality of a dataset~\cite{PCA:1901}. The approach taken in this paper to computing the PCA of a matrix $\mat A$ relies on the SVD of the upper-triangular matrix $\mat R$ in the QR decomposition of $\mat A$. The novelty of \Figaro over all prior approaches to PCA is that it does not require the materialization of $\mat A$ to compute $\mat R$, in case $\mat A$ is defined by database joins. A further approach to PCA over database joins was recently proposed in the database theory literature, without a supporting implementation~\cite{SparseTensor:TODS:2020}: It uses the min-max theorem based on the Rayleigh quotient~\cite{Strang:LAbook:2006} and computes iteratively one eigenvector of $\mat A^\transpose\mat A$ at a time. The benefit of that approach is that the computation of $\mat A^\transpose\mat A$ can be pushed past joins and may take time less than materializing the matrix $\mat A$ representing the join result.

\section{Conclusion}

This article introduces \Figaro, an algorithm that computes the QR decomposition of matrices defined by joins over relational data. By pushing the computation past the joins, \Figaro can significantly reduce the number of computation steps and rounding errors. \Figaro can also be used to compute singular value and eigenvalue decompositions as well as the principal component analysis of the non-materialized matrix representing the joins over relational databases. We experimentally verified that \Figaro can outperform in runtime and in many cases also in accuracy the linear algebra package Intel MKL.
In the future, we plan to extend \Figaro to work with categorical data directly, to avoid one-hot encoding such data.
 
\begin{acknowledgements}
This project has received funding
from the {European Union's Horizon 2020 research and innovation
programme} under grant agreement No {682588}.
\end{acknowledgements}

\bibliographystyle{plain}
\bibliography{refs}  %

\appendix

\section{Further Details on the Accuracy Experiment}
\label{sec:accuracy-design}

We show that given an upper-triangular matrix $\mat R_\text{fixed}$, we can devise matrices $\mat S$ and $\mat T$ with arbitrary dimensions such that $\mat S \times \mat T = \mat Q \mat R$ for an upper-triangular $\mat R = \begin{bmatrix}
\mat R_\text{fixed} & \mat V \\
\mat 0 & \mat W 
\end{bmatrix}$, for some matrices $\mat V$, $\mat W$.

We denote by $\mat{1}_{m \times n}$ the $m \times n$ matrix that consists entirely of $1$s and by $\mat{0}_{m \times n}$ the $m\times n$ matrix that consists entirely of $0$s. 

We revisit the notion of Kronecker product. For an $m \times n$ matrix $\mat A$ and a $p \times q$ matrix $\mat B$, the Kronecker product $\mat A \otimes \mat B$ is the $mp \times nq$ matrix
\[\begin{bmatrix}
\mat A[1:1] \mat B & \cdots & A[1:n] \mat B \\
\vdots & \ddots & \vdots \\
\mat A[m:1] \mat B & \cdots & A[m:n] \mat B
\end{bmatrix},\] where each $A[i:j] \mat B$ is the matrix $\mat B$ multiplied by the scalar $A[i:j]$. 
We can express a Cartesian product in terms of Kronecker products: for $\mat{S} \in \matspace{m_1}{n_1}$ and $\mat{T} \in \matspace{m_2}{n_2}$, we have $\mat S \times \mat T = \begin{bmatrix}
  \mat{S} \otimes \mat{1}_{m_2 \times 1} & \mat{1}_{m_1 \times 1}  \otimes \mat{T}
\end{bmatrix}$.

We use the following two observations. 

\begin{lemma}[\cite{van2000ubiquitous}]  \label{lemma:QRKronProd}
For arbitrary real matrices $\mat{A}$ and $\mat B$ with respective QR decompositions $\mat{A} = \mat{Q}_A \mat{R}_A$ and $\mat{B} = \mat{Q}_B \mat{R}_B$ it holds $\mat{A} \otimes \mat{B} = (\mat{Q}_A \otimes \mat{Q}_B)(\mat{R}_A \otimes \mat{R}_B)$.
\end{lemma}

\begin{lemma}
Let $\mat{A} \in \matspace{m}{n_1}$ and $\mat B \in \matspace{m}{n_2}$ be arbitrary and let $\mat{A} = \mat{Q}_A \mat{R}_A$ be the QR decomposition of $\mat A$, where $\mat{Q}_A \in \matspace{m}{n_1}, \mat{R}_A \in \matspace{n_1}{n_1}$. There is an orthogonal matrix $\mat Q' \in \matspace{m}{n_2}$ and matrices $\mat{V} \in \matspace{n_1}{n_2}, \mat{W} \in \matspace{n_2}{n_2}$ such that
     \[\begin{bmatrix}
          \mat{A} & \mat{B}
      \end{bmatrix}
      =
      \begin{bmatrix}
          \mat{Q}_A & \mat{Q}'
      \end{bmatrix}
      \begin{bmatrix}
          \mat{R}_A & \mat{V}\\
          \mat{0}_{n_2 \times n_1} & \mat{W}
      \end{bmatrix}. \]
  \label{lemma:QRHorCat}
\end{lemma}

From Lemma~\ref{lemma:QRKronProd} it follows that for a matrix $\mat{A} \in \matspace{m}{n}$ with QR decomposition $\mat A = \mat Q \mat R$, we have 
\begin{align*}
\mat{A} \otimes \mat{1}_{m \times 1} & = (\mat{Q} \otimes \mat{Q}_1) (\mat{R} \sqrt{m}) \\
 \mat{1}_{m \times 1} \otimes  \mat{A} & = (\mat{Q}_1 \otimes  \mat{Q}) (\mat{R} \sqrt{m}),
\end{align*}
where $\mat{1}_{m \times 1} = \mat{Q}_1 \begin{bmatrix}
\sqrt{m}
\end{bmatrix}$ is the QR decomposition of  $\mat{1}_{m \times 1}$, $\begin{bmatrix}
\sqrt{m}
\end{bmatrix}$ is a $1\times 1$ matrix with only entry $\sqrt m$, and $\mat{R} \sqrt{m}$ is the multiplication of the matrix $\mat R$ with the scalar $\sqrt m$.

Putting the above observations together, we obtain:
\begin{corollary}
    Let $\mat{S} \in \matspace{m_1}{n_1}, \mat{T} \in \matspace{m_2}{n_2}$ be arbitrary and let $\mat{S} = \mat{Q}_S \mat{R}_S$ be the QR decomposition of $\mat S$, where $\mat{Q}_S \in \matspace{m_1}{n_1}, \mat{R}_S \in \matspace{n_1}{n_1}$. There is an orthogonal matrix $\mat Q' \in \matspace{m_1 m_2}{n_2}$ and matrices $\mat{V} \in \matspace{n_1}{n_2}, \mat{W} \in \matspace{n_2}{n_2}$ such that
\begin{align*}
\mat S \times \mat T & =
        \begin{bmatrix}
          \mat{S}  \otimes \mat{1}_{m_2 \times 1} & \mat{1}_{m_1 \times 1}  \otimes \mat{T}
        \end{bmatrix} \\
       & =
        \begin{bmatrix}
            \mat{Q}_S \otimes \mat{Q}_1 & \mat{Q}'
        \end{bmatrix}
        \begin{bmatrix}
            \mat{R}_S \sqrt{m_2} & \mat{V}\\
            \mat{0}_{n_2 \times n_1} & \mat{W}
        \end{bmatrix}.
\end{align*}  
    \label{cor:CartProd}
\end{corollary}

Let $\mat R_S \in \matspace{n_1}{n_1}$ be an arbitrary upper-triangular matrix. We arbitrarily choose a vector $\mat {v} = (v_1, \ldots, v_{m_1})^\transpose \in \mathbb{Q}^{m_1}$ with $\norm{\mat v} = 1$ and a $m_2 \times n_2$ matrix $\mat T$ of natural numbers, where $m_2$ is square, so $\sqrt{m_2}$ is a natural number. 
We set $\mat Q_S$ to be the first $n_1$ columns of the orthogonal matrix of rational numbers \cite{zihwei2006extending}
\[\mat{\hat Q} = \begin{bmatrix}
     v_1 & v_2 & v_3 & \cdots & v_{n_1} \\
     v_2 & \frac{v^2_2-v_1 - 1}{v_1 + 1} & \frac{v_2 v_3}{v_1 + 1} & \cdots & \frac{v_2 v_{m_1}}{v_1 + 1} \\
     v_3 & \frac{v_3 v_2}{v_1 + 1} & \frac{v_3^2 - v_1 - 1}{v_1 +  1} & \cdots & \frac{v_3 v_{m_1}}{v_1 + 1} \\
     \vdots & \vdots & \vdots & \ddots & \vdots \\
     v_{m_1} & \frac{v_{m_1} v_2}{v_1 + 1} & \frac{v_{m_1} v_3}{v_1 + 1}& \cdots & \frac{v_{m_1}^2 - v_1 - 1}{v_1 + 1}
 \end{bmatrix},\]
so  $\mat Q_S = \mat{\hat Q}[: \{1, \ldots, n_1\}]$.
We obtain $\mat S$ as $\mat S = \mat Q_S \mat R_S$.

It follows from Corollary~\ref{cor:CartProd} that there is an orthogonal matrix $\mat Q$ as well as matrices $\mat V, \mat W$ such that $\mat S \times \mat T = \mat Q \begin{bmatrix}
\mat R_S \sqrt{m_2} & \mat V \\
\mat 0 & \mat W 
\end{bmatrix}$, as desired. Furthermore, if $\mat R_S$ only consists of rational numbers then so do $\mat S$, $\mat T$ and $\mat R_S \sqrt{m_2}$.

When computing the QR decomposition of $\mat S \times \mat T$ we can compare the ground truth $\mat R_S \sqrt{m_2}$ with the corresponding part of the computed result.

\end{document}